\documentclass[final,5p,times,twocolumn]{elsarticle}

\usepackage{lineno,hyperref}

\usepackage{subfigure}
\usepackage{color} 
\usepackage{amsmath}
\usepackage{amsfonts,amssymb} 
\usepackage{multirow}
\usepackage{array}
\usepackage{booktabs}
\usepackage{nicematrix}
\usepackage{stfloats}
\usepackage{lipsum}
\usepackage{cuted}
\usepackage{subfigure}

\modulolinenumbers[5]

\journal{Journal of Elsevier}

\biboptions{sort&compress}

\begin{document}

\begin{frontmatter}

\title{Cross-Modality Fusion Transformer for Multispectral Object Detection}


\author{Fang Qingyun}
\ead{fqy17@mails.tsinghua.edu.cn}
\author{Han Dapeng}
\ead{dphan@mail.tsinghua.edu.cn}
\author{Wang Zhaokui\corref{cor1}}
\cortext[cor1]{Corresponding author}
\ead{wangzk@tsinghua.edu.cn}
\address{School of Aerospace Engineering, Tsinghua University, Beijing, 100084 China}

%
%
%

%
%
%
%

\begin{abstract}

Multispectral image pairs can provide combined information, making object detection applications more reliable and robust in the open world.
To fully exploit the different modalities, a simple yet effective cross-modality feature fusion approach, named Cross-Modality Fusion Transformer (CFT) is presented in this paper.
Unlike prior CNNs-based works, our network learns long-range dependencies and integrates global contextual information in the feature extraction stage under the guidance of the Transformer scheme.
More importantly, by leveraging self attention of the Transformer, the network can naturally execute intra-modality and inter-modality fusion simultaneously, and robustly capture the latent interactions between RGB and thermal domains.
Thereby the performance of multispectral object detection is improved significantly.
Extensive experiments and ablation studies on multiple datasets demonstrate that the proposed scheme is effective and obtains state-of-the-art detection performance.
Our code and models are  available at \href{https://github.com/DocF/multispectral-object-detection}{https://github.com/DocF/multispectral-object-detection}.
\end{abstract}


\end{frontmatter}


\section{Introduction}
In real-world object detection applications,  the environment is often open and dynamic,  requiring models and algorithms to deal with the challenges caused by openness, such as rain, fog, occlusions, poor lighting, low resolution, etc. 
It is difficult for an algorithm to use only visible-band sensor data to achieve high accuracy under these conditions.
Hence, the multispectral imaging technology is being adopted, given its ability to provide the combined information coming from multispectral cameras e.g., visible and thermal.
By fusing the complementarity of different modalities, the perceptibility, reliability, and robustness of the detection algorithms can be improved.

Recent advances in convolutional neural networks (CNNs), more specifically,  the invention of two-stream CNN-based detectors, have increased detection performance in  the field of multispectral object detection\cite{MultispectralLiu2016,ParkUnified2018,LiMultispectral2018,LiIllumination2018,ZhangCross2019,ZhangMultispectral2020,tiong2020multimodal, ZhangGuided2021,ChenMultimodal2021,SharmaYOLOrs2021,qingyun2022cross}.
In addition, some challenging multispectral datasets, e.g.,  FLIR\cite{Freeflir}, LLVIP\cite{JiaLLVIP2021}, VEDAI\cite{RazakarivonyVehicle2016}, have also driven the advancement of this technology.

\begin{figure}
	\centering
	\includegraphics[width=0.48\columnwidth]{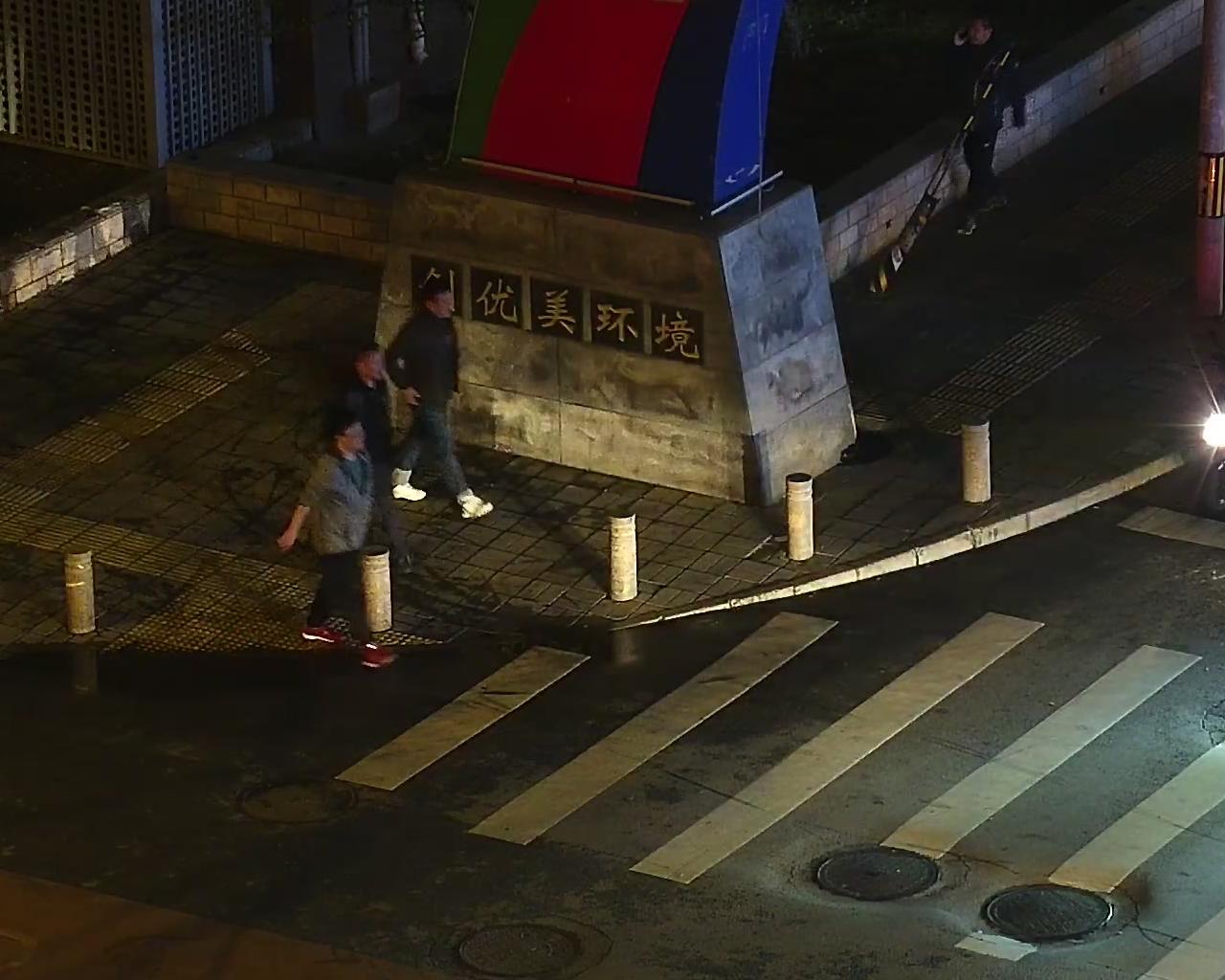}\hspace{0.075cm}
	\includegraphics[width=0.48\columnwidth]{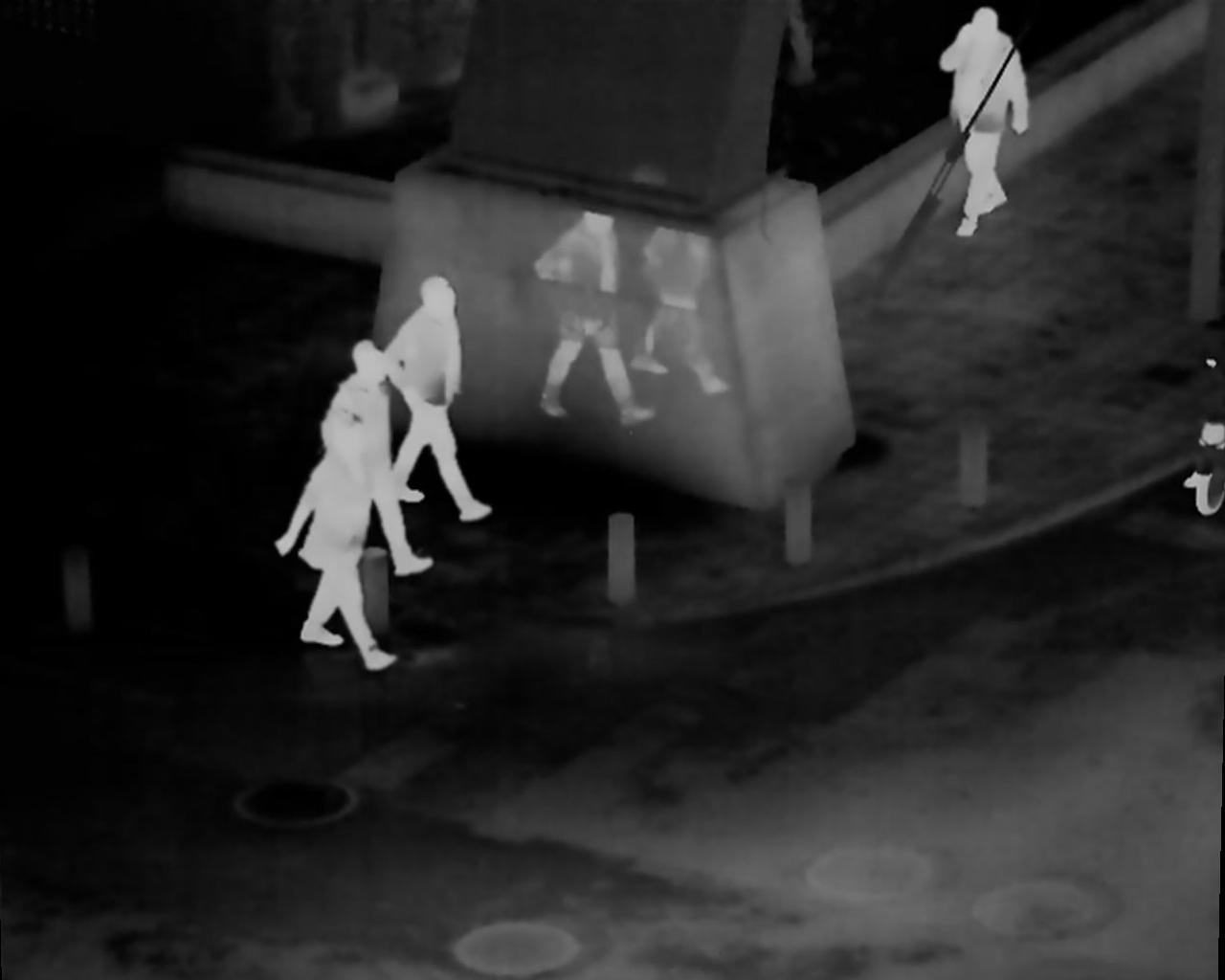} \\
	\vspace{0.10cm}
	\includegraphics[width=0.48\columnwidth]{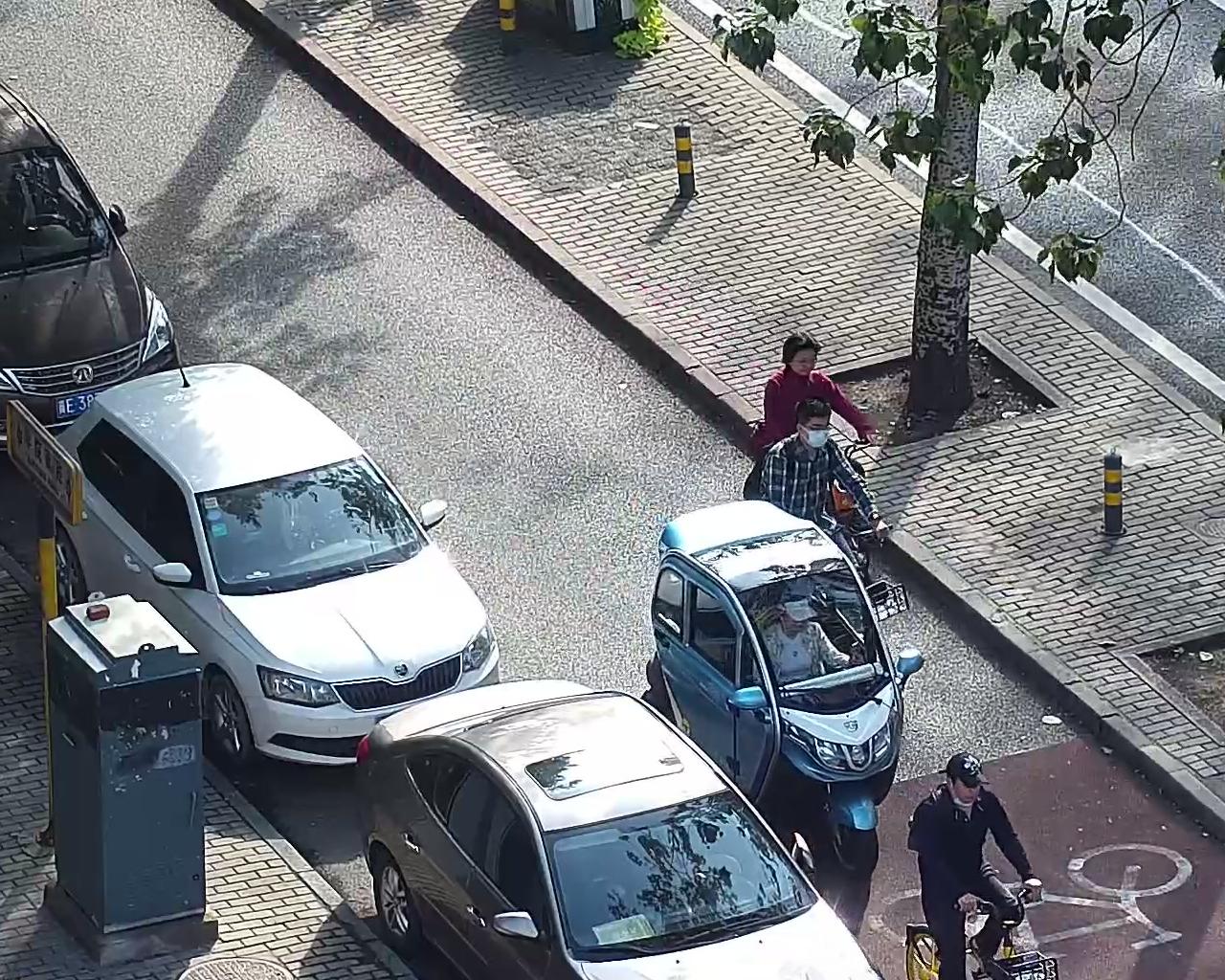}\hspace{0.075cm}
	\includegraphics[width=0.48\columnwidth]{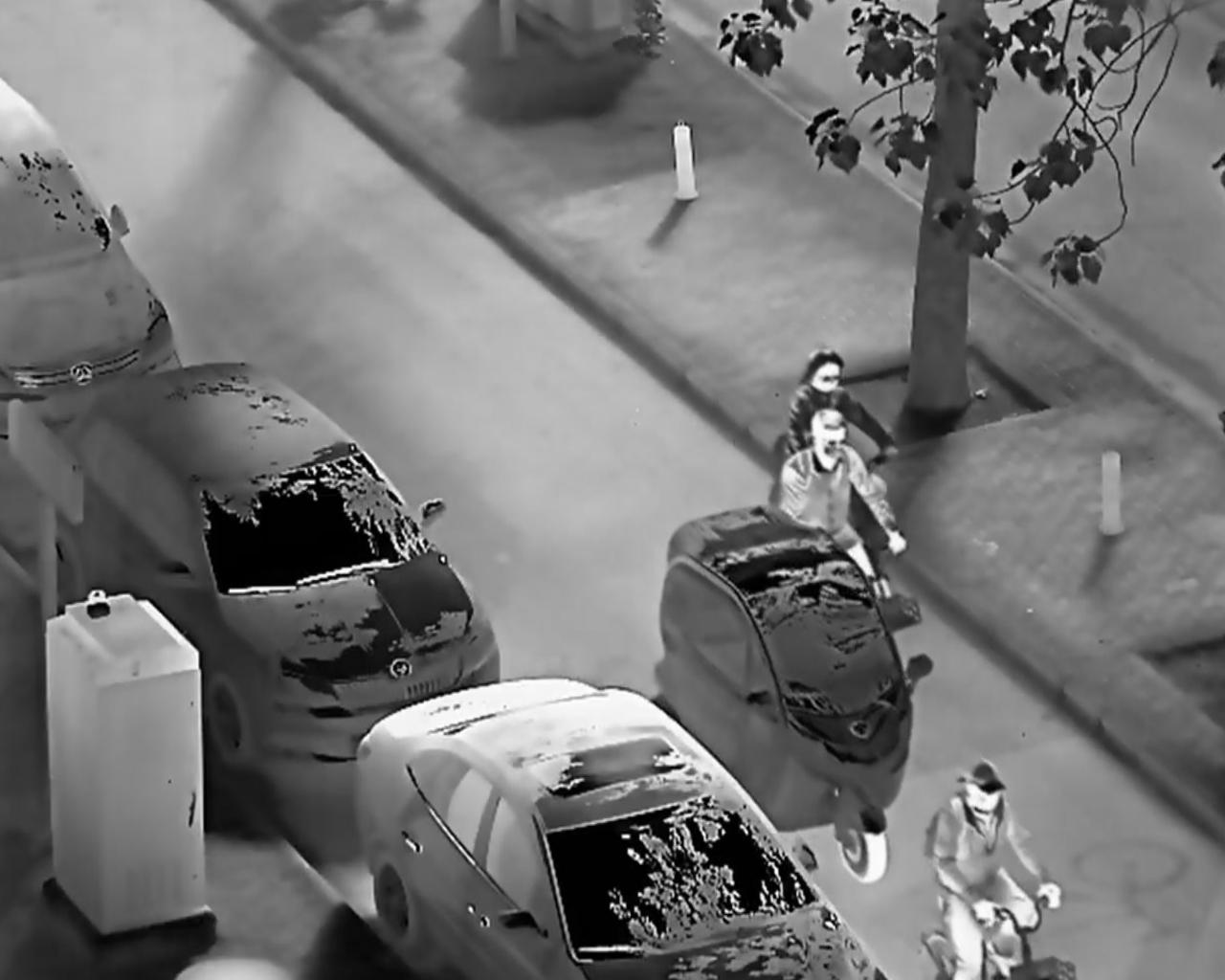}
	
	\caption{Visible-infrared paired examples from LLVIP.
		The paired images in the first row are  captured in nighttime traffic scenes.  
		Compared with the visual image on the left, the thermal image on the right can  capture a clearer contours of pedestrians under insufficient illumination conditions. 
		Besides, the thermal image also captures pedestrians obscured by a pillar.
		The paired images in the second row are  captured in daytime.
		During well-lit daytime, the visual image has more details, such as edges, textures, and colors, than thermal images. 
		With these details, we can easily find the driver hidden in the motor three wheeler. 
		However, the driver is difficult to be found in the thermal image.
		Zoom in to see details.
	}
	\label{fig:examples} 
\end{figure}

Figure~\ref{fig:examples} illustrates the advantages of multispectral images over visible-only or thermal-only images under different conditions.  
However, the exploitation of multispectral data will raise new challenges:
How to integrate the representations to fully take advantage of the inherent complementarity between different modalities?
And how to design an effective cross-modality fusion mechanism for maximum performance gain?

In the prior works, no matter how to design the modal fusion mechanism, they are all mostly based on deep convolutional neural networks\cite{MultispectralLiu2016, LiMultispectral2018, ZhangCross2019, ZhangGuided2021, ChenMultimodal2021}.
There is considerable literature that has proven that CNNs can have strong representation learning capabilities within a single intra-modal reasoning \cite{KrizhevskyAlexnet2012, HeResnet2015, GirshickRCNN2014, RenFaster2017, Liussd2016, Redmonyolo2016, qingyun2020efficient, amudhan2022lightweight}, especially for visual modality.
However,  it is non-trivial to extend them to cross-modality fusion or modality interaction to fully exploit the inherent complementarity.
The convolution operator of CNNs can be described as a non-fully connected graph in which each spatial position in the feature maps is consider as a node. 
Since the convolution operator has a non-global receptive field, the information is only integrated in a local area.
In contrast, the self attention operator of Transformer \cite{VaswaniAttention2017} can be regarded as a fully connected graph, so it can learn long-range dependencies and its receptive field can be global. 
Therefore, the self attention of Transformer provides a natural mechanism which is more suitable for connecting multimodal signals than the convolution of CNNs.
{  Multimodal Transformers have been applied to various tasks including image segmentation \cite{Linwei2019Cross, Efthymio2021Into}, cross-modal sequence generation\cite{Jiaman2020Learning,Ruilong2021Learn}, video retrieval\cite{Valentin2020Multi,Maksim2021MDMMT} and image/video captioning/classification \cite{Jiasen2019ViLBERT,Chen2019VideoBERT,XusongBERT4SessRec2019,Guang2019Entangled,Vladimir2020Multi}.
However, there is no work which designs the Transformer for multispectral object detection in the literature.}

We propose a  novel and effective multispectral fusion approach, which is called  \textit{Cross-Modality Fusion Transformer} (CFT), to explore the potential of Transformer in the application of multispectral object detection.
Specifically, our CFT modules are embedded in the feature extraction backbone in order to integrate global contextual information from different modalities.
To the best of our knowledge, this is the first work adopting the Transformer for multispectral object detection.

{
	
	Contributions: (1) We introduce a new and powerful two-stream backbone that enhances one modality from another modality under the guidance of the Transformer scheme.
	(2) We propose a simple yet effective CFT module, and give theoretical insights into it, showing that the CFT module simultaneously fuses the intra-modality and inter-modality features.
	(3) Considerable experiments show that the current method achieves state-of-the-art performance on three public datasets.
}

\section{Related work}

Fusing the features of two modalities is the core problem in multispectral object detection, which can be described as
\begin{equation}\label{eq_fusion}
	\mathbf{F}_{Fused} = \mathcal{F}\left(\mathbf{F}_{R},\mathbf{F}_{T}\right) =  \mathcal{F}\left(\phi_R\left(\mathbf{I}_{R}\right),\phi_T\left(\mathbf{I}_{T}\right)\right).
\end{equation}
$\mathbf{I}_{R}$ and $\mathbf{I}_{T}$ represents the input RGB image and the input Thermal image, respectively.
$\mathbf{F}_{R}$ denotes the RGB feature maps
and $\mathbf{F}_{T}$ denotes  the Thermal feature maps.
The feature extraction functions of networks, $\phi_R(.)$ and $\phi_T(.)$, are applied to generate the feature maps for the different modal input images.
$\mathbf{F}_{Fused}$ indicates the fused feature maps, and $\mathcal{F(.)}$ is the function of fusion.
In the Eq.~\eqref{eq_fusion}, multi-modality fusion can be divided into two aspects: input variable (i.e., input features), and fusion functions.
Following the idea of Eq.~\eqref{eq_fusion}, the previously published investigations in the literature can also be sorted into two categories, one is focued on the input features (``macro" level), and the other is focued on constructing the fusion functions (``micro" level).

\textbf{Macro level.} 
On this level, researchers aim to solve the problem of where to fuse, that is, which stage of the input features to choose to fuse.
Most of them explore the best fusion stage by designing the macro network architectures.
The first study of this type \cite{WagnerMultispectral2016} investigates two deep fusion architectures (early fusion and late fusion) and analyzes their performance on multispectral data.
To explore the potential of deep models for multispectral pedestrian detection further, Liu et al. \cite{MultispectralLiu2016} designed another two ConvNet fusion architectures (halfway fusion and score fusion) and demonstrated that the halfway fusion model achieved the best detection synergy.
Since then, several subsequent works\cite{LiMultispectral2018, LiIllumination2018,   ZhangGuided2021, ZhangWeakly2019}  illustrated that the halfway fusion overwhelmingly outperforms the other three fusion architectures.

\textbf{Mirco level.} 
Besides the macro network architectures, the construction of fusion function is another key point for complementary learning of modality interaction.
Naturally,  the most straightforward way is to utilize concatenation, element-wise addition, element-wise average/maximum, and element-wise cross product to merge feature maps of RGB and thermal modalities directly.
Two variations of novel Gated Fusion Units (GFU) \cite{ZhengGFD2019} are proposed in GFD-SSD to investigate the combination of feature maps generated by the two SSD middle layers. 
Zhang et. al. \cite{ZhangMultispectral2020} proposed a novel cycle fuse-and-refine module to improve the multispectral feature fusion while achieving the complementary/consistency balance of the features. 
Transformer\cite{VaswaniAttention2017} is originally a classic Natural Language Processing (NLP) model proposed by the Google team in 2017.
However, due to its strong representation ability and concise model, it has been extended to computer vision and multimodal fields in recent years \cite{Linwei2019Cross, Efthymio2021Into,Jiaman2020Learning,Ruilong2021Learn,Jiasen2019ViLBERT,Chen2019VideoBERT,XusongBERT4SessRec2019,Guang2019Entangled,Vladimir2020Multi,Valentin2020Multi,Maksim2021MDMMT}.
Different from the prior methods, a Transformer-based scheme is proposed to fuse intra-modal and inter-modal information for multispectral object detection in current paper.
%

\section{Methodology}

\begin{figure*}[htbp]
	\centerline{\includegraphics[width=1\linewidth]{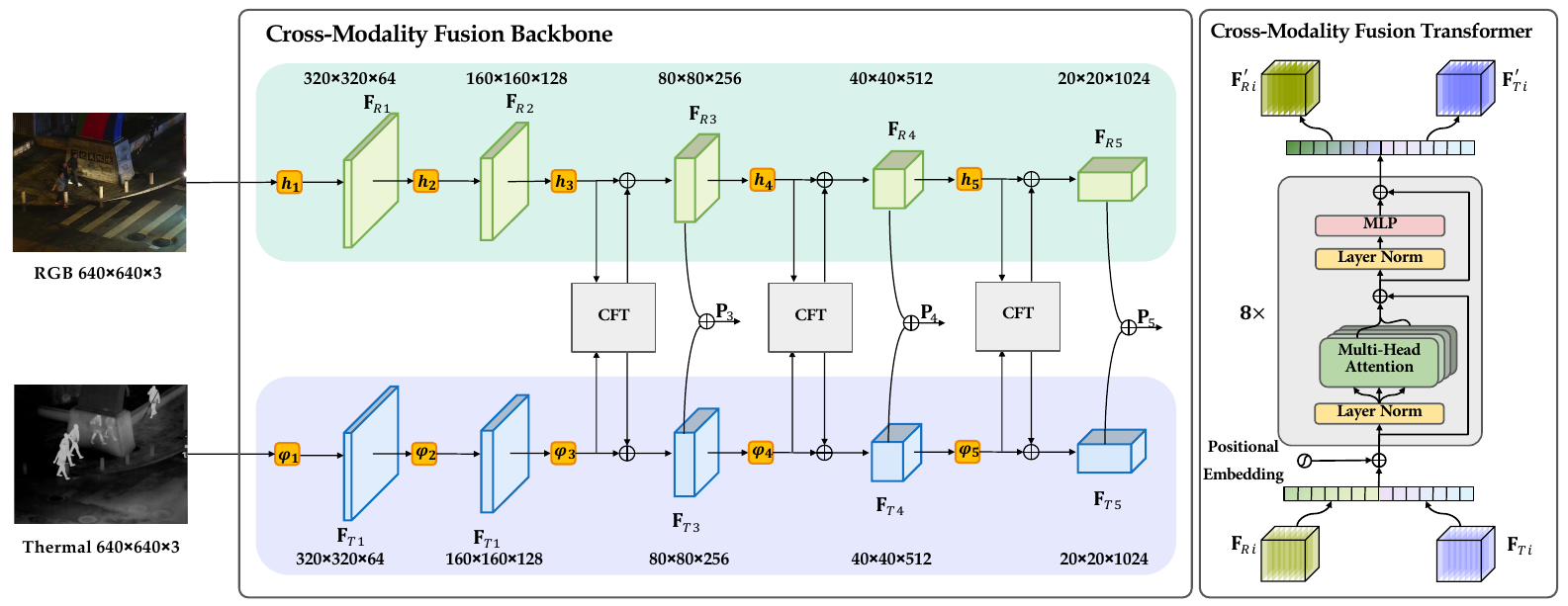}}
	\caption{ 
		Framework of Cross-Modality Fusion Backbone.
		The backbone has two parts: a two-stream feature extraction network and three Cross-Modality Fusion Transformer modules.
		Among them, $h_i$ and $\varphi_i$ are the convolution modules of the RGB and thermal branches, $\mathbf{F}_{Ri}$ and $\mathbf{F}_{Ti}$ are the feature maps of their respective modalities, and $\mathbf{P}_i$ represents the input of the subsequent feature pyramid. 
		The right side shows the design details of the Cross-Modality Fusion Transformer module.
	}
	\label{fig_cft}
\end{figure*}

\textbf{Overview.}
To demonstrate the effectiveness of our proposed CFT fusion module, we extend the framework of YOLOv5, to enable multispectral object detection.
To be precise, we redesign the YOLOv5 feature extraction network as a two-stream backbone, which is similar to GFD-SSD and embedded the CFT modules to facilitate modal fusion and modal interaction, named as \textit{Cross-Modality Fusion Backbone} (CFB).
An illustration of our \textit{Cross-Modality Fusion Backbone} and \textit{Cross-Modality Fusion Transformer}
is presented in Fig.~\ref{fig_cft}.

\textbf{Details.} 
Specifically, given the intermediate RGB convolution feature maps $\mathbf{F}_{R}\in \mathbb{R}^{C\times H\times W}$, and thermal convolution feature maps $\mathbf{F}_{T}\in \mathbb{R}^{C\times H\times W}$,
the sentences $\mathbf{I}_{R}\in \mathbb{R}^{HW \times C}$ and  $\mathbf{I}_{T}\in \mathbb{R}^{HW \times C}$ are obtained by flattening each feature map and permuting the order of the matrices.
Second, we concatenate the sentences of each modality and add a learnable positional embedding, which is a trainable parameter of dimension $2HW\times C$, to get the input sentences  $\mathbf{I}\in \mathbb{R}^{2HW \times C}$ of the Transformer.
The positional embedding enables the model to differentiate spatial information  between different tokens at train time\cite{VaswaniAttention2017}. 
Third, the input sequence $\mathbf{I}$ is projected onto three weight matrices to compute a set of queries, keys and values ($\mathbf{Q}$, $\mathbf{K}$ and $\mathbf{V}$),
\begin{align}
	\mathbf{Q}&=\mathbf{I} \mathbf{W}^{Q},\\
	\mathbf{K}&= \mathbf{I} \mathbf{W}^{K},\\
	\mathbf{V}&=\mathbf{I} \mathbf{W}^{V},
\end{align}
where $\mathbf{W}^{Q} \in \mathbb{R}^{C\times D_Q}$, $\mathbf{W}^{K} \in \mathbb{R}^{C\times D_K}$ and $\mathbf{W}^{V} \in \mathbb{R}^{C\times D_V}$ are weight matrices.  
Moreover, $D_Q$, $D_K$ and $D_V$ are equal in our Transformer , i.e., $D_Q=D_K=D_V=C$.
Fourth, the self attention layer uses the scaled dot products between $\mathbf{Q}$ and $\mathbf{K}$ to compute the attention weights and then multiply by the values to infer the refined output $\mathbf{Z}$,
\begin{equation}
	\mathbf{Z}=\operatorname{Attention}(\mathbf{Q}, \mathbf{K}, \mathbf{V}) = \operatorname{softmax}\left(\frac{\mathbf{Q K}^{T}}{\sqrt{D_K}}\right) \mathbf{V}, 
\end{equation}
where $\frac{1}{\sqrt{D_K}}$ is a scaling factor for preventing the softmax function from falling into a region with extremely small gradients when the magnitude of dot products grow large.
To encapsulate multiple complex relationships from different representation subspaces at different positions, the multi-head attention mechanism is adopted,
\begin{equation}
	\begin{aligned}
		& \mathbf{Z'} = \operatorname{MultiHead}(\mathbf{Q}, \mathbf{K}, \mathbf{V}) =\operatorname{Concat}\left(\mathbf{Z}_{1}, \dots, 	\mathbf{Z}_{h} \right) \mathbf{W}^{O}, \\
		& \mathbf{Z}_{i} =\operatorname{Attention}\left( \mathbf{Q} \mathbf{W}_{i}^{Q}, \mathbf{K} \mathbf{W}_{i}^{K}, \mathbf{V} \mathbf{W}_{i}^{V}\right).
	\end{aligned}
\end{equation}
The subscript $h$ denotes the number of heads, and $\mathbf{W}^{O} \in \mathbb{R}^{h\cdot C \times C  }$ denotes the projected matrix of  $\operatorname{Concat}\left(\mathbf{Z}_{1}, \dots, \mathbf{Z}_{h}\right)$.
Then the Transformer uses a two-layer fully connected feed-forward network with a GELU \cite{hendrycks2016gaussian} activation in between to calculate the output sequences $\mathbf{O}$, which are of the same shape as input sequences $\mathbf{I}$,
\begin{align}
	\mathbf{O} &=\operatorname{MLP}(\mathbf{Z''})+\mathbf{Z''}, \\
	& = \operatorname{FC}_{2}\left(\operatorname{GELU}\left( \operatorname{FC}_{1}(\mathbf{Z''})\right)\right)+ \mathbf{Z''}
\end{align}
where $\mathbf{Z''} = \mathbf{Z'} + \mathbf{I}$. 
Finally, exploiting the inverse operation of the first step, the output sentences $\mathbf{O}$ are converted into the recalibration results $\mathbf{F}_{R}'$ and  $\mathbf{F}_{T}'$ and added to the original modality branch as complementary information.

\textbf{Implementations.}
The parameter amount and computational complexity of a standard Transformer block can be formalized as 
\begin{align}
\label{params}	\text{Params}\sim O(\text{Transformer})=4HWC+8C^{2},\\
 \label{flops}  \text{FLOPs}\sim\Omega(\text{Transformer})=12HWC^{2}+2(HW)^{2}C.
\end{align}
 A CFT module has 8 duplicate Transformer blocks, as shown in Fig.~\ref{fig_cft}.
In addition, since the dimension of the input sentences  $\mathbf{I}$  is $2HW\times C$, the actual expression of $HW$  in the above Eq.~\eqref{params} and \eqref{flops}  is $2HW$.
Apart from the Parameters and the Floating Point Operations (FLOPs) , the memory access also needs to be considered \cite{MaShufflenet2018}, especially when calculating the dot product of queries and keys, an intermediate matrix of $2HW\times2HW$ dimensions will be generated.
When the input picture size is $640 \times 640$, after two downsamplings ($H=W=160$), the elements of the matrix $\mathbf{Q K}^{T}$ exceed 2.4G, which is unacceptable for ordinary computers.

{For reducing expensive computation, our solution is to use a global average pooling 
that downsamples the feature maps to a low and fixed resolution of $H = W = 8$ before passing them to the Transformer block.
And the output is upsampled by bilinear interpolation to the original resolution before being added to the original mode branch.
In this way, the number of parameters and computational complexity of our multispectral detector are acceptable (cf. the Parameters and FLOPs in Table \ref{tab_params_flops}).

As mentioned before, YOLOv5 is chosen to be our basic detector. 
By adding an additional branch for the extraction of thermal features, it is transformed into a two-stream convolutional neural network, which constitutes the baseline.
In other words, the difference between the baseline and our transformer-based fusion detection algorithm is only the CFT module.}

\begin{figure}[htbp]
	\centerline{\includegraphics[width=0.95\linewidth]{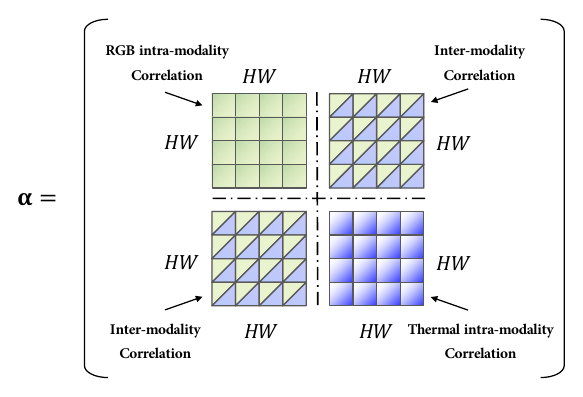}}
	\caption{ Illustration of the correlation matrix $\boldsymbol{\alpha}$.
	}
	\label{fig_trans_matrix}
\end{figure}

\textbf{Why Transformer?}
The main idea behind our module is leveraging the self attention mechanism to learn the binary relationship of  RGB and thermal modalities, more precisely, leveraging the correlation matrix to weight each position of the input feature maps.
It can be formulated as Eq.\eqref{eq_alpha}. In the formula, ${\alpha}_{i,j}$ represents the correlation between the $i$-th position and the $j$-th position on the feature maps.
\color{black} According to the Eq. \eqref{eq_alpha},  four matrix blocks can be inferred naturally, when calculating the correlation matrix $\boldsymbol{\alpha}$.
Two of them are intra-modality correlation matrix blocks (RGB and thermal), and the other two are inter-modality correlation matrix blocks, as illustrated in Fig. \ref{fig_trans_matrix}.

\begin{figure*}
	\color{black}
	\NiceMatrixOptions{code-for-first-row = \color{red},
		code-for-first-col = \color{blue}}
	\begin{equation}\label{eq_alpha}
		\boldsymbol{\alpha}  = \operatorname{softmax}\left(\frac{\mathbf{Q K}^{T}}{\sqrt{D_K}}\right)  =
		\begin{pNiceArray}{cccc|cccc}[first-row,first-col,nullify-dots]
			& C_1 &  &  \Cdots & C_{HW} &  C_{HW+1} &   & \Cdots &   C_{2HW}  \\
			L_1 & \alpha_{1,1} & \alpha_{1,2} & \cdots & \alpha_{1,HW} &   \alpha_{1,HW+1}  & \alpha_{1,HW+2}& \cdots & \alpha_{1,2HW}  \\
			& \alpha_{2,1} & \alpha_{2,2} & \cdots & \alpha_{2,HW} &   \alpha_{2,HW+1}  & \alpha_{2,HW+2}& \cdots & \alpha_{2,2HW}  \\
			\Vdots & \vdots & \vdots & \ddots & \vdots &  \vdots  & \vdots & \ddots & \vdots  \\
			L_{HW} & \alpha_{HW,1} & \alpha_{HW,2} & \cdots & \alpha_{HW,HW} &   \alpha_{HW,HW+1}  & \alpha_{HW,HW+2}& \cdots & \alpha_{HW,2HW}  \\
			\hline
			L_{HW+1}	& \alpha_{HW+1,1} & \alpha_{HW+1,2} & \cdots & \alpha_{HW+1,HW} &   \alpha_{HW+1,HW+1}  & \alpha_{HW+1,HW+2}& \cdots & \alpha_{HW+1,2HW}  \\
			& \alpha_{HW+2,1} & \alpha_{HW+2,2} & \cdots & \alpha_{HW+2,HW} &   \alpha_{HW+2,HW+1}  & \alpha_{HW+2,HW+2}& \cdots & \alpha_{HW+2,2HW}  \\
			\Vdots & \vdots & \vdots & \ddots & \vdots &  \vdots  & \vdots & \ddots & \vdots  \\
			L_{2HW}	 & \alpha_{2HW,1} & \alpha_{2HW,2} & \cdots & \alpha_{2HW,HW} &   \alpha_{2HW,HW+1}  & \alpha_{2HW,HW+2}& \cdots & \alpha_{2HW,2HW}  \\
		\end{pNiceArray},
	\end{equation}
\end{figure*}

Hence, with the help of Transformer, we don't need to carefully design the modal fusion module.
We just need to simply splice the multi-modal features into a sequence, and then the Transformer can automatically perform simultaneous intra-modality and inter-modality information fusion  and robustly capture the latent interactions between RGB and hermal domains.

%

\section{Experiment}

\subsection{Datasets}
All experments are evaluated on three benchmark datasets, i.e, FLIR\cite{Freeflir} , LLVIP\cite{JiaLLVIP2021}  and VEDAI\cite{RazakarivonyVehicle2016}.

\textbf{FLIR.}
The FLIR ADAS dataset is a challenging multispectral object detection dataset that includes day and night scenes.
There are a lot of unaligned image pairs in the original data set, which makes network training difficult.
Therefore, an ``aligned" version \cite{ZhangMultispectral2020} is recently released that manually removes unaligned visible-thermal image pairs.
This new dataset contains 5,142 well-aligned multispectral image pairs, of which 4,129 pairs are used for training and 1,013 pairs are used for testing and cover three object categories: ``person", ``car" and ``bicycle".
{We conduct experiments on the "aligned" version of the FLIR dataset in current work, and for convenience, the FLIR that appears later all refer to its "aligned" version.}

\textbf{LLVIP.}
LLVIP is a very recently released visible-infrared paired pedestrians dataset for low-light vision. 
This dataset contains 33672 images, or 16836 pairs, most of which were taken in low-light environments, and all of the images are strictly spatio-temporal aligned. 

\textbf{VEDAI.}
In addition to the above two ground-view datasets, we also test our method on VEDAI  which is a multispectral aerial imagery dataset for vehicle detection.
The dataset contains nine vehicle classes for a total of more than 3700 annotated targets in more than 1200 images in two different resolutions (1024 × 1024 and 512 × 512). 
Both RGB and IR modalities are available for each image.

\subsection{Experimental Settings}

\subsubsection{Loss Function}
In this section, the loss functions utilized for training our proposed mutlispectral object detectors is introduced.
Formally, the overall loss function is a sum of the bounding-box regression loss ($\mathcal{L}_{\text{box}}$), the classification loss($\mathcal{L}_{\text{cls}}$) and the confidence loss ($\mathcal{L}_{\text {conf}}$),
\begin{equation}
	\begin{aligned}
		\mathcal{L}_{\text {total }}&=\mathcal{L}_{\text {box }}+\mathcal{L}_{\text {cls }}+\mathcal{L}_{\text {conf }}\\
		&=\mathcal{L}_{\text {box }}+\mathcal{L}_{\text {cls }}+\mathcal{L}_{\text {noobj }}+\mathcal{L}_{\text {obj}}, 
	\end{aligned}
\end{equation}
where,
\begin{align}
	&\begin{aligned}\label{lossbox}
		\mathcal{L}_{\text {box}} &= \sum_{i=0}^{S^2} \sum_{j=0}^N  \mathbf{1}_{i, j}^{o b j} \cdot 	\mathcal{L}_{\text {GIoU}_i}  \\
		&= \sum_{i=0}^{S^2} \sum_{j=0}^N   \mathbf{1}_{i, j}^{obj} \cdot\left[ {1 - \text {GIoU}_i}\right]  \\
		& = \sum_{i=0}^{S^2} \sum_{j=0}^N   \mathbf{1}_{i, j}^{obj} \cdot \left[ 1- \frac{B^g_i \cap B^p_i}{B^g_i \cup B^p_i} +\frac{B^c_i \backslash(B^g_i \cup B^p_i)}{B^c_i} \right],
	\end{aligned} \\
	&\begin{aligned}
		\mathcal{L}_{\text {cls}}=\sum_{i=0}^{S^2} \sum_{j=0}^N  \mathbf{1}_{i, j}^{obj} \cdot \sum_{c \in \text { classes }} p_i(c) \log \left(\hat{p}_i(c)\right),
	\end{aligned}\\
	&\begin{aligned}\label{lossnoobj}
	\mathcal{L}_{\mathrm{noobj}} &=\sum_{i=0}^{S^2} \sum_{j=0}^N \mathbf{1}_{i, j}^{n o o b j} \cdot\left(c_i-\hat{c}_i\right)^2, 
	\end{aligned}\\
	&\begin{aligned}
	\mathcal{L}_{\mathrm{obj}} &=\sum_{i=0}^{S^2} \sum_{j=0}^N \mathbf{1}_{i, j}^{obj} \cdot\left(c_i-\hat{c}_i\right)^2.
	\end{aligned}
\end{align}
As Eq.~\eqref{lossbox} shows, Generalized Intersection over Union (GIoU) loss\cite{rezatofighi2019generalized} is employed as the predicted box regression loss.
GIoU loss can be a better choice compared to IoU loss, no matter which IoU-based performance measure is ultimately used.
${S^2}$ and $N$ denote the number of image grids during prediction and the number of the predicted boxes in each grid.
$B^g$, $B^p$, and $B^c$ are the ground truth, the predicted box, and the smallest enclosing box surrounding $B^g$ and $B^p$, respectively.
The coefficient $\mathbf{1}_{i, j}^{obj}$ indicates whether the $j$th predicted box of the $i$th grid is a positive sample.
The classification loss $\mathcal{L}_{\text {cls}}$ takes the cross-entropy form, $p(c)$ represents the probability that the real sample is class $c$, and $\hat{p}(c)$ represents the probability that the network predicts the sample to be class $c$.
The last confidence loss consists of two components, $\mathcal{L}_{\text {noobj}}$ and $\mathcal{L}_{\text{obj}}$, both of which are squared-error losses.
The coefficient $\mathbf{1}_{i,j}^{noobj}$ in Eq.~\eqref{lossnoobj} has the opposite definition to the previous coefficient $\mathbf{1}_{i, j}^{obj}$.
Finally, $c$ and $\hat{c}$ is represents the true value of the confidence and the predicted confidence by the network.

\subsubsection{Training Details}
We use an stochastic gradient descent (SGD) optimizer with an initial learning rate of 1e-2, a momentum of 0.937, and a weight decay of 0.0005. 
{All models are trained on two NVIDIA® TITAN RTX™ GPUs for 200 epochs with a batch size of 32. 
In pursuit of better performance, we adopt the YOLOV5 model pre-trained on the COCO dataset \cite{lin2014microsoft} as weight initialization.}
As for data augmentation, we use the Mosaic method \cite{Alexey2020yolov4} which mixes four training images in one image.

\subsection{Evaluation Metrics}
All models are evaluated with three object detection metrics introduced with MS-COCO: mean Average Precision (mAP), mAP50 and mAP75.

\begin{equation}\label{equ_map}
	\text {mAP}=\frac{1}{n}\sum_{i=0}^{n} \text{AP}_i = \frac{1}{n}\sum_{i=0}^{n}\int_{0}^{1} P_i(r) ~\mathrm{d} r
\end{equation}
where,\begin{equation}\label{equ_ap}
	\begin{aligned}
			\text{AP}_i & = \int_{0}^{1} P_i(r) \text{ d} r \\
		& =\int_{0}^{1}\text { Precision } \text{d} (\text{Recall}) \\
		& = \int_{0}^{1} \frac{\text{TP}}{\text{TP}+\text{FP}} \text{d} \frac{\text{TP}}{\text{TP}+\text{FN}}
	\end{aligned}
\end{equation}
TP means true positive, which is that a predicted box by detectors and the ground truth (GT) meet the intersection over union (IoU) threshold;
otherwise, it will be considered as a false positive (FP). 
False negative (FN) means there is a true target, but the detector doesn't find it. 
Equation~\eqref{equ_ap} indicates that $\text{AP}$ is the integral of the Precision-Recall Curve (PRC) for each category. 
mAP50 computes the mean of all the AP values for all categories at IoU=0.50 in the Eq.~\eqref{equ_map}.
Similarly, mAP75 calculate the mean at  IoU=0.75.
mAP is the primary challenge metric, which can be formulated as the mean at IoU=0.50:0.05:0.95. 
Obviously, it is much stricter than the other two metrics.

\subsection{Ablation Study}
In Table \ref{tab_ab},  the detection performances on different datasets (FLIR, LLVIP, and VEDAI) are compared.
The best records and the improvements are marked in bold and by {\color{blue}$\uparrow$}, respectively.
{The mAP values of the RGB-only and thermal-only YOLOv5 on the FLIR dataset are 31.8\% and 39.5\% respectively, and the thermal-only detection result surpasses the RGB-only, which also appears in the LLVIP dataset.
However, on the VEDAI dataset, the RGB-only YOLOv5 performs slightly better than the thermal-only one (mAP: {\color{blue}$\uparrow$0.1}). 
It is possible that  there are a lot of low-light scenes in FLIR and LLVIP, resulting in the loss of effective target areas.
Furthermore, comparing the mAP of the thermal-only YOLOv5 and the two-stream baseline, the simple two-stream network cannot fully exploit the inherent complementarity between different modalities. 
What is more, these coarse approaches may increase the difficulty of network learning and aggravate the imbalance of the modalities, which results in performance degradation.  
It can be observed in Table \ref{tab_ab} that, with our proposed CFT, the performance of the detector has improved on all three datasets. 
Especially for the VEDAI dataset, the evaluation metric mAP75 is increased by 18.2\%, and mAP is elevated by 9.2\%.

\begin{table*}[htbp]

	\centering
	\caption{Comparisons of  performances with different datasets in terms of  mAP50, mAP75, and mAP.}
	\setlength{\tabcolsep}{3pt}
	\begin{tabular}{p{1.5cm}|p{2.0cm}<{\centering}|p{2cm}<{\centering}|p{1.8cm}<{\centering}|p{1.8cm}<{\centering}|p{1.8cm}<{\centering}}
		\toprule
		Dataset & Modality & Method  &   mAP50 & mAP75 & mAP\\
		\midrule
		\multirow{4}{*}{FLIR} &RGB & YOLOV5  & 67.8 & 25.9 & 31.8 \\
									& Thermal & YOLOV5   & 73.9&\textbf{35.7} &39.5\\
	 								& RGB+T  & + Two Stream  &73.0 & 32.0 & 37.4\\
									& RGB+T  & + CFT & \textbf{78.7} ({\color{blue}$\uparrow$5.7}) & {35.5} ({\color{blue}$\uparrow$3.5})  & \textbf{40.2} ({\color{blue}$\uparrow$2.8})  \\   
		\midrule
		
		\multirow{4}{*}{LLVIP}   & RGB & YOLOV5   &90.8 & 51.9 & 50.0\\
										  & Thermal & YOLOV5 & 94.6 & {72.2 } &61.9\\
										  & RGB+T & + Two Stream & 95.8 &71.4 & 62.3 \\
										 & RGB+T &+ CFT  &\textbf{97.5} ({\color{blue}$\uparrow$1.7})  & \textbf{72.9} ({\color{blue}$\uparrow$1.5})   &\textbf{63.6} ({\color{blue}$\uparrow$1.3})  \\
		
		\midrule
		
		\multirow{3}{*}{VEDAI}  &RGB  &  YOLOV5   & 74.3 & 46.9 & 46.2\\
										  &Thermal  &  YOLOV5  &  74.0 & 46.8 & 46.1\\
										  & RGB+T & + Two Stream &  79.7 & 47.7  & 46.8 \\
										  & RGB+T &+ CFT   & \textbf{85.3} ({\color{blue}$\uparrow$5.6}) &\textbf{65.9} ({\color{blue}$\uparrow$18.2}) &\textbf{56.0} ({\color{blue}$\uparrow$9.2}) \\   
		\bottomrule
	\end{tabular}
	\label{tab_ab}
\end{table*}

To present the general effectiveness of our CFT, we further combine the CFT module with other classical detectors, i.e., YOLOv3\cite{redmon2018yolov3} and Faster R-CNN\cite{RenFaster2017}, and test them on the FLIR dataset.
The results of the experiments are evaluated in Table \ref{tab_params_flops}.
Evaluation metrics include both efficiency (i.e., network parameters and GFLOPs) and effectiveness (i.e., mAP50, mAP75, and mAP).
As shown in Table \ref{tab_params_flops}, the proposed CFT module  improves the performance of multispectral object detection using either the one-stage or two-stage detector by a clear margin.
Specifically,  the CFT approach achieves a 5.7\% gain on mAP50, a 3.5\% increment of mAP75,
and a 2.8\%  advancement over mAP (on YOLOV5). 
When combing with YOLOV3, the performance gains are 4.0\%, 1.4\%, and 2.2\%, respectively.
When adopting Faster R-CNN as the basic detector, our CFT outperforms the baseline by 4.3\%, 2.6\%, 2.1\% in terms of mAP50, mAP75, and mAP, respectively.  

\begin{table*}[htbp]

	\centering
	\caption{Comparisons of performances on the FLIR dataset in terms of network parameters (Param.), Giga Floating Point Operations (GFLOPs), mAP50, mAP75, and mAP. }
	\setlength{\tabcolsep}{3pt}
	\begin{tabular}{p{1.8cm}|p{2.5cm}<{\centering}|p{2.2cm}<{\centering}|p{2cm}<{\centering}|p{2cm}<{\centering}|p{2cm}<{\centering}|p{2cm}<{\centering}|p{2cm}<{\centering}}
		\toprule
		Modality & Method & Detector &  Param.  &GFLOPs  & mAP50 & mAP75 & mAP \\
		\midrule
		RGB & CSPDarknet53 & \multirow{4}{*}{YOLOV5} &   47.06M &115.58 &  67.8 & 25.9 & 31.8 \\
		Thermal& CSPDarknet53 &  									 &   47.06M &115.58 &  73.9&\textbf{35.7} &39.5\\
		RGB+T & + Two Stream & 										&  73.72M  & 190.14 &  73.0 & 32.0 & 37.4\\  
		RGB+T & + CFT & 												&   206.03M &224.40 &  \textbf{78.7} ({\color{blue}$\uparrow$5.7}) & {35.5} ({\color{blue}$\uparrow$3.5})  & \textbf{40.2} ({\color{blue}$\uparrow$2.8}) \\
		\midrule
		RGB & Darknet53 & \multirow{3}{*}{YOLOV3}    &  61.50M &154.67 &   58.3 & 19.8 &25.7 \\
		Thermal & Darknet53 & 								   &  61.50M &154.67 &  73.6 & 31.3 & 36.8 \\
		RGB+T & + Two Stream &  								& 102.12M &256.46&  72.3  & 30.6  & 36.1 \\  
		RGB+T & + CFT & 											& 234.44M &290.73  &  \textbf{76.3} ({\color{blue}$\uparrow$4.0})  & \textbf{32.0} ({\color{blue}$\uparrow$1.4}) & \textbf{38.3} ({\color{blue}$\uparrow$2.2})  \\   
		\midrule
		RGB & Resnet53 & \multirow{4}{*}{Faster R-CNN}&   41.13M  & 75.59 &64.9  & 21.1 &28.9   \\
		Thermal & Resnet53 & &   41.13M  & 75.59  & 74.4 & 32.5 &  37.6 \\   
		RGB+T & + Two Stream & & 64.61M & 102.50   & 73.1 & 32.0    &37.1 \\  
		RGB+T & + CFT & & 196.93M & 136.77  & \textbf{77.5} ({\color{blue}$\uparrow$4.3}) & \textbf{34.6} ({\color{blue}$\uparrow$2.6})  & \textbf{39.2} ({\color{blue}$\uparrow$2.1}) \\
		\bottomrule
	\end{tabular}
	\label{tab_params_flops}
\end{table*}

}

\begin{figure}[htbp]
	\centering
	{\includegraphics[width=0.46\linewidth]{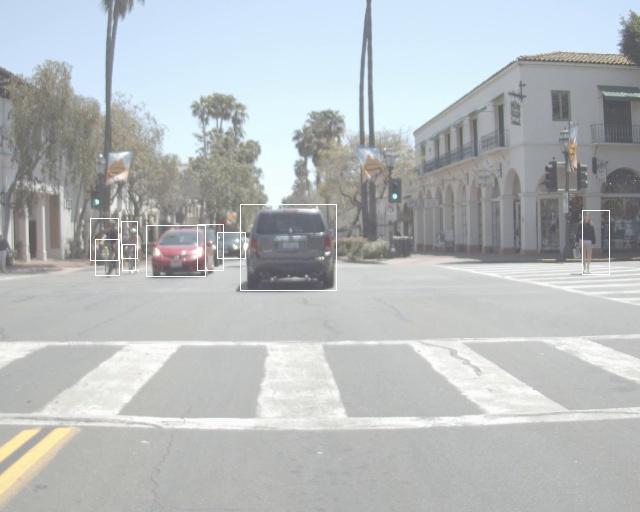}} \hspace{0.05cm}
	{\includegraphics[width=0.46\linewidth]{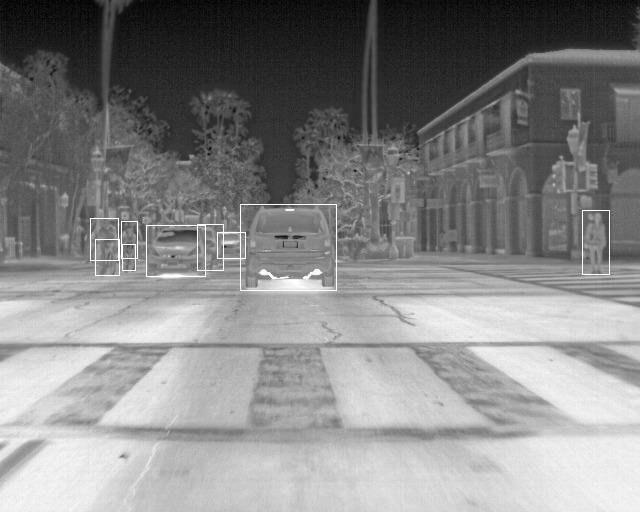}} \\
	\vspace{0.1cm}
	{\includegraphics[width=0.46\linewidth]{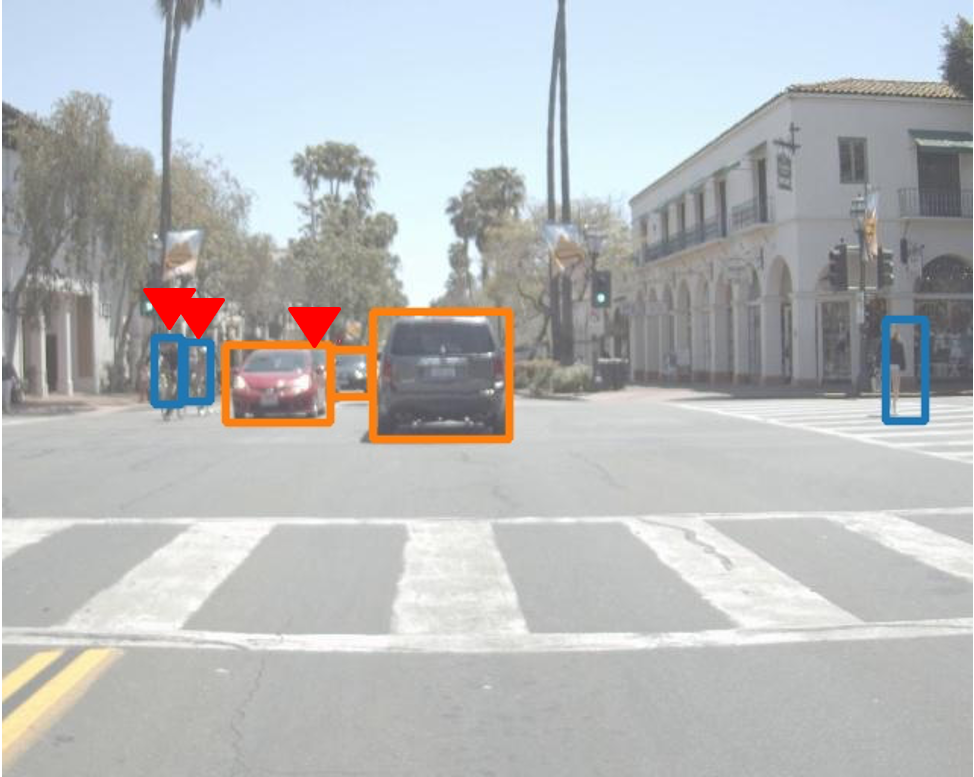}} \hspace{0.05cm}
	{\includegraphics[width=0.46\linewidth]{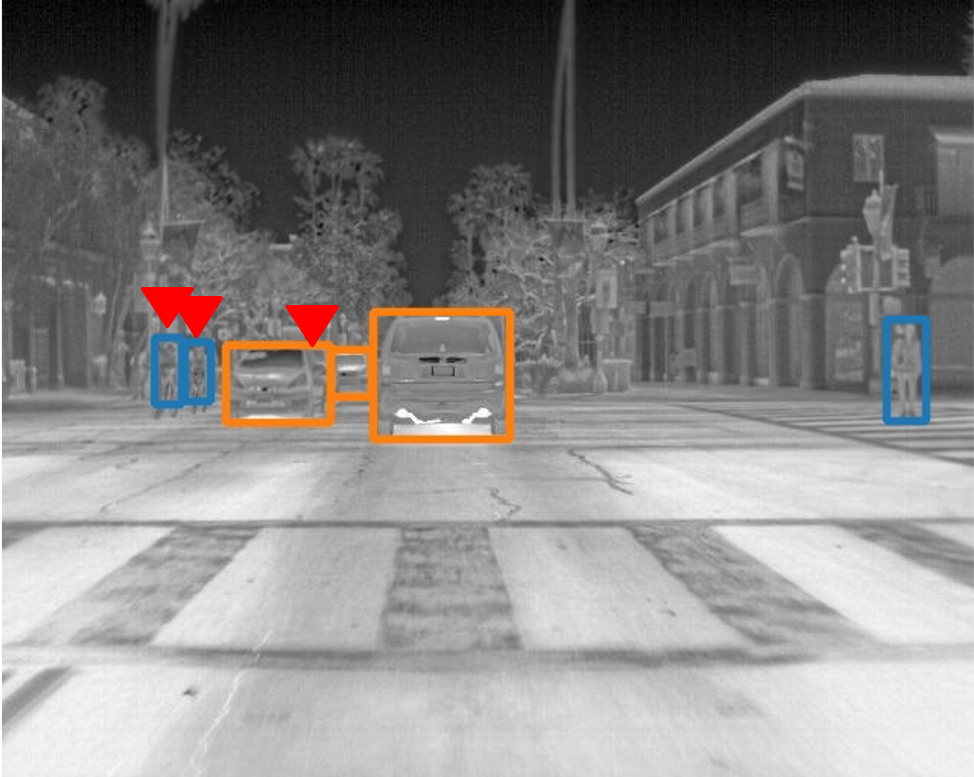}} \\
	\vspace{0.1cm}
	{\includegraphics[width=0.46\linewidth]{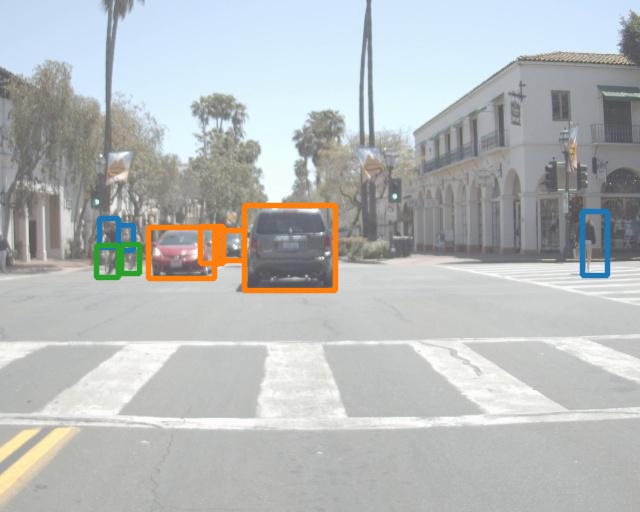}} \hspace{0.05cm}
	{\includegraphics[width=0.46\linewidth]{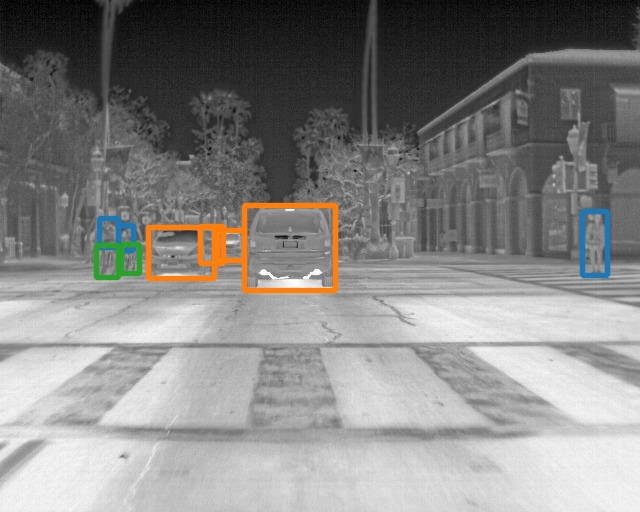}}
	\caption{ 
		Qualitative comparison of multispectral object detection in the FLIR ADAS dataset. 
		First column: color images, second column: thermal images. 
		From top row to bottom row: ground truth, detection results of the baseline, detection results of our method. 
		Note that the {\color{red} red inverted triangles} indicate FNs. 
		Zoom in for more detail.
	}
	\label{fig_example_flir}
\end{figure}

\begin{figure}[htbp]
	\centering
	{\includegraphics[width=0.46\linewidth]{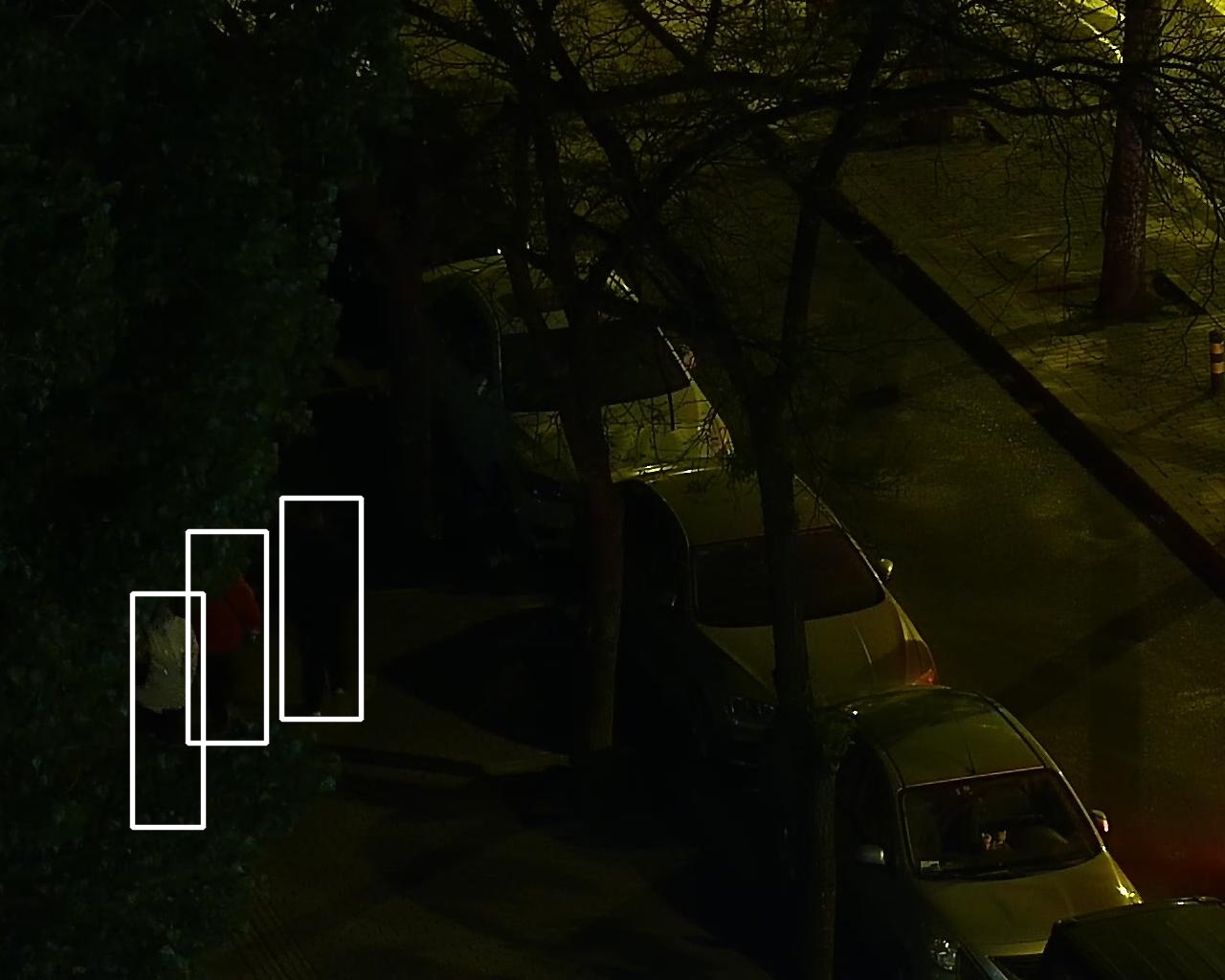}} \hspace{0.05cm}
	{\includegraphics[width=0.46\linewidth]{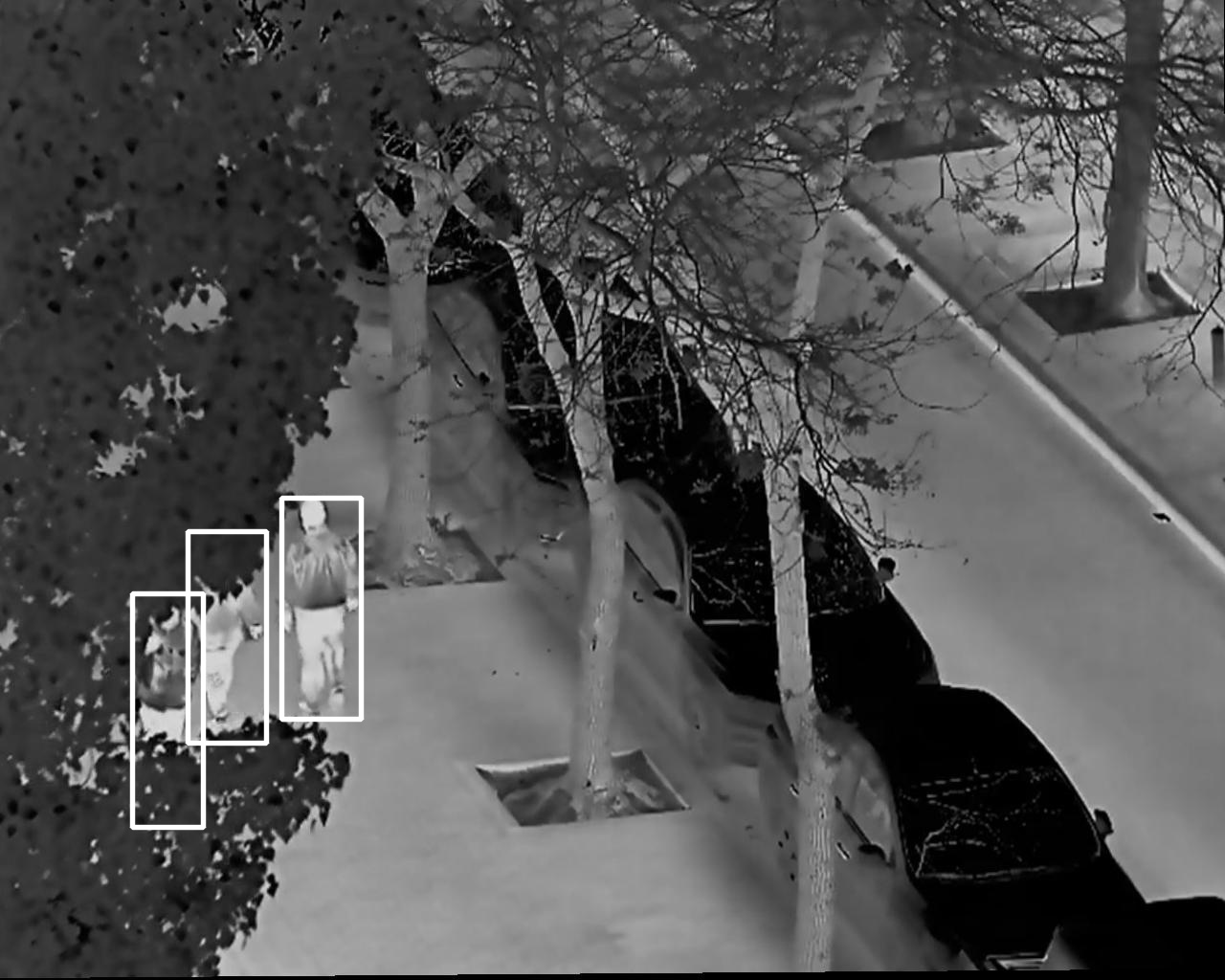}} \\
	\vspace{0.1cm}
	{\includegraphics[width=0.46\linewidth]{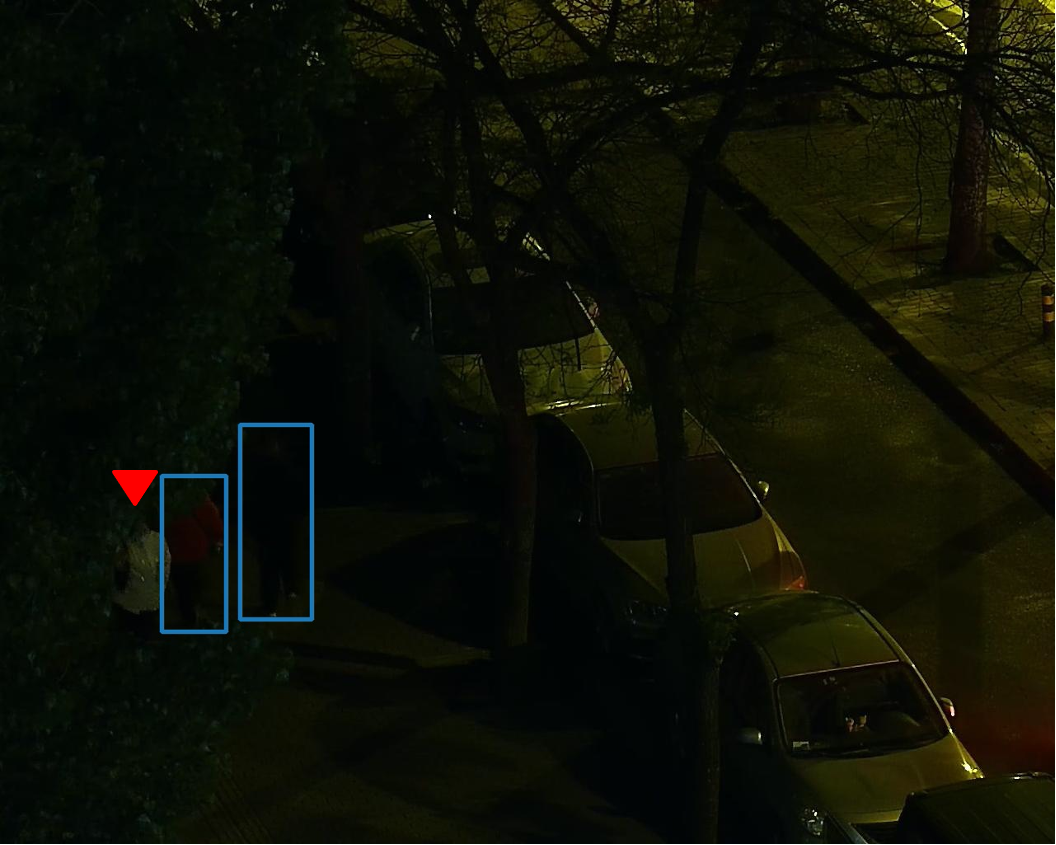}} \hspace{0.05cm}
	{\includegraphics[width=0.46\linewidth]{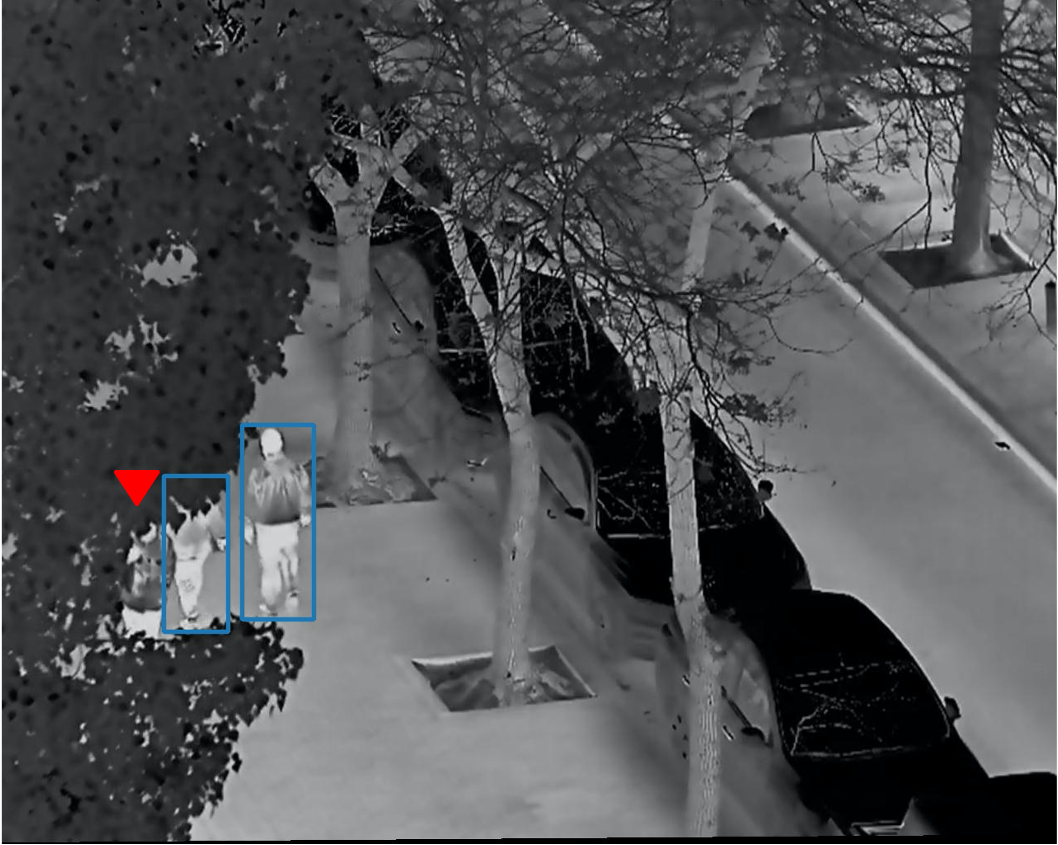}} \\
	\vspace{0.1cm}
	{\includegraphics[width=0.46\linewidth]{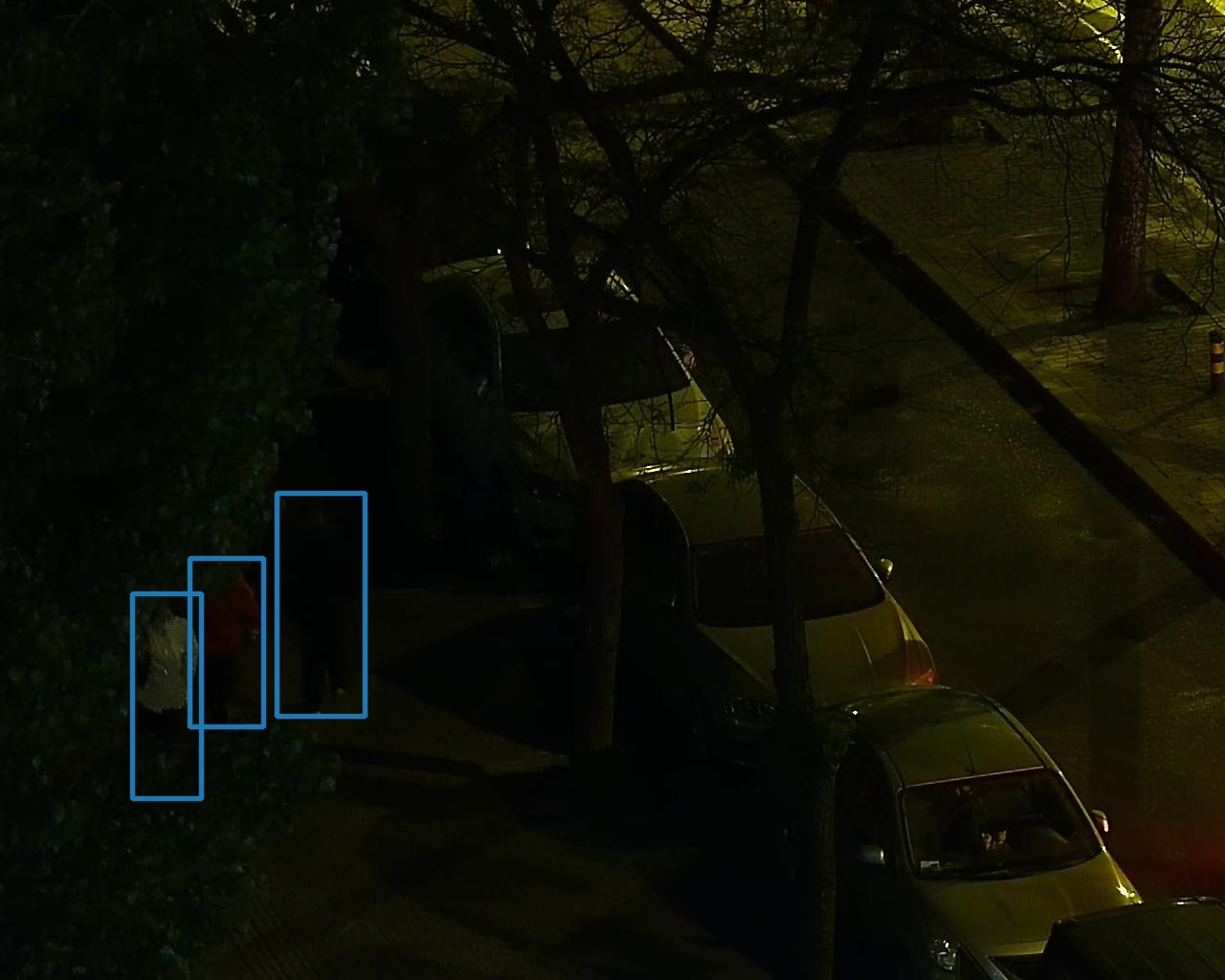}} \hspace{0.05cm}
	{\includegraphics[width=0.46\linewidth]{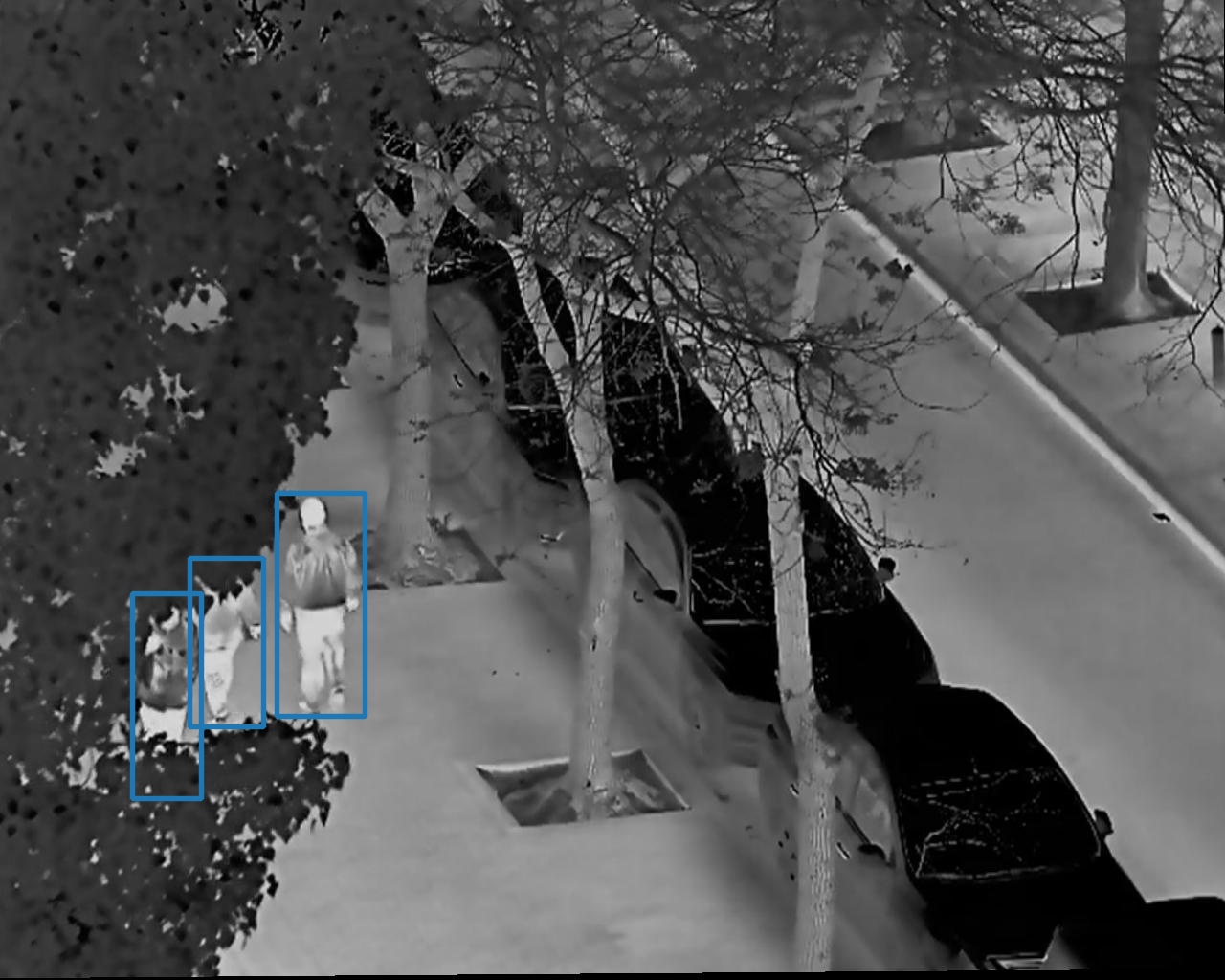}}
	\caption{
		Qualitative comparison of multispectral object detection in the LLVIP dataset. 
		The arrangement of the figure is the same as in Fig.~\ref{fig_example_flir}.
		Note that {\color{red} red inverted triangles} indicate FNs. 
		Zoom in for more detail.
	}
	\label{fig_example_llvip}
\end{figure}

\begin{figure}[htbp]
	\centering
	{\includegraphics[width=0.46\linewidth]{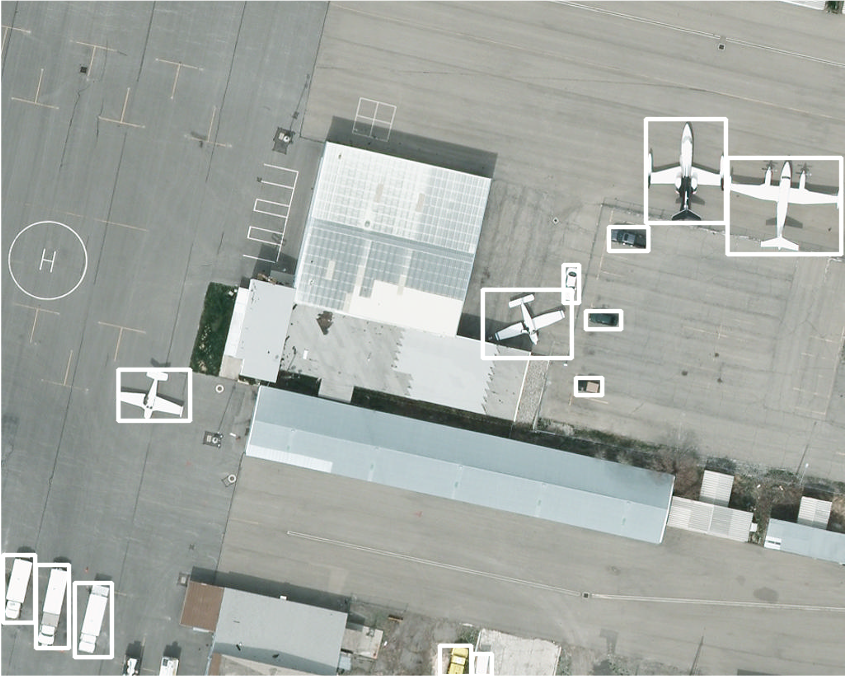}} \hspace{0.05cm}
	{\includegraphics[width=0.46\linewidth]{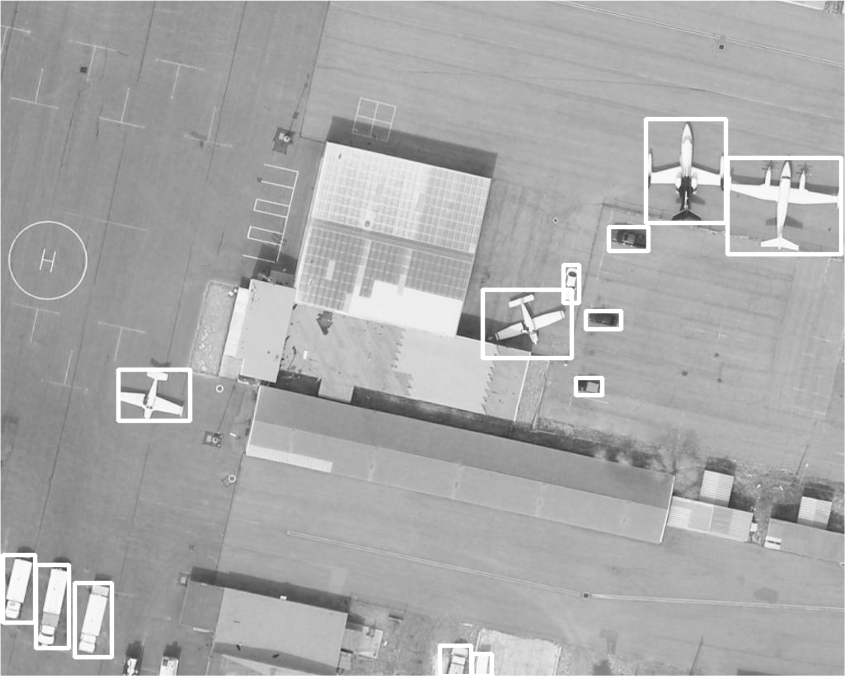}} \\
	\vspace{0.1cm}
	{\includegraphics[width=0.46\linewidth]{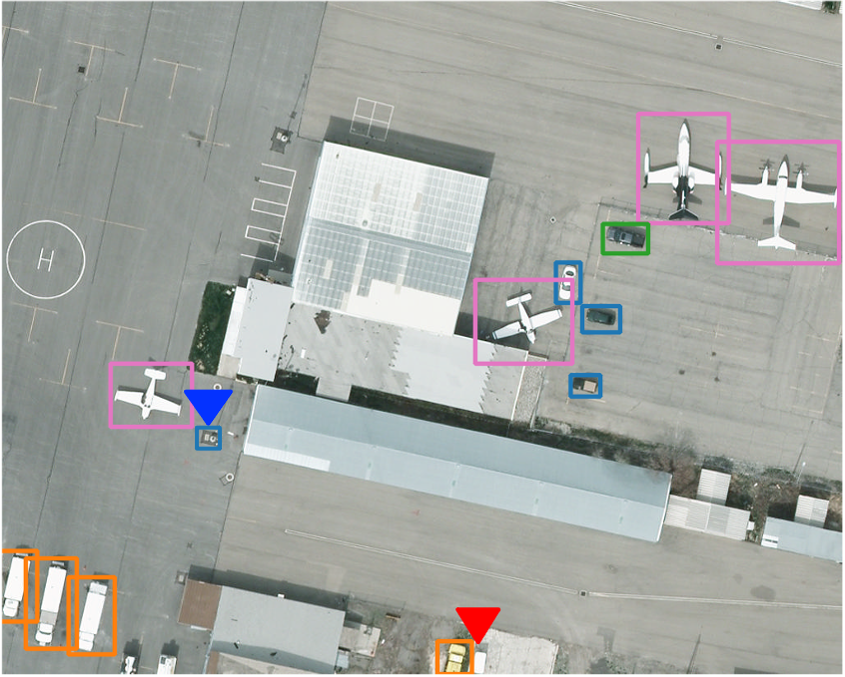}} \hspace{0.05cm}
	{\includegraphics[width=0.46\linewidth]{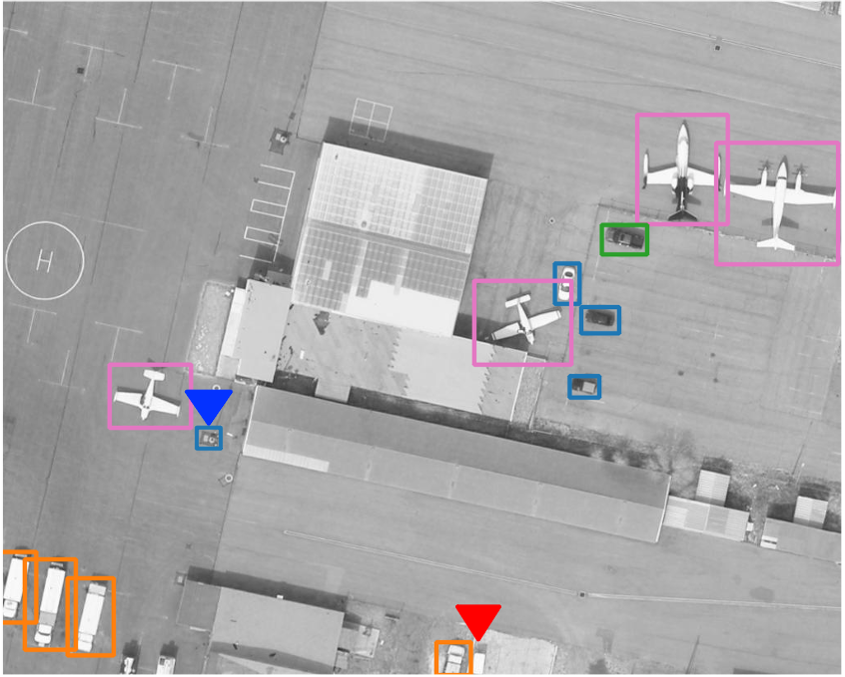}}\\
	\vspace{0.1cm}
	{\includegraphics[width=0.46\linewidth]{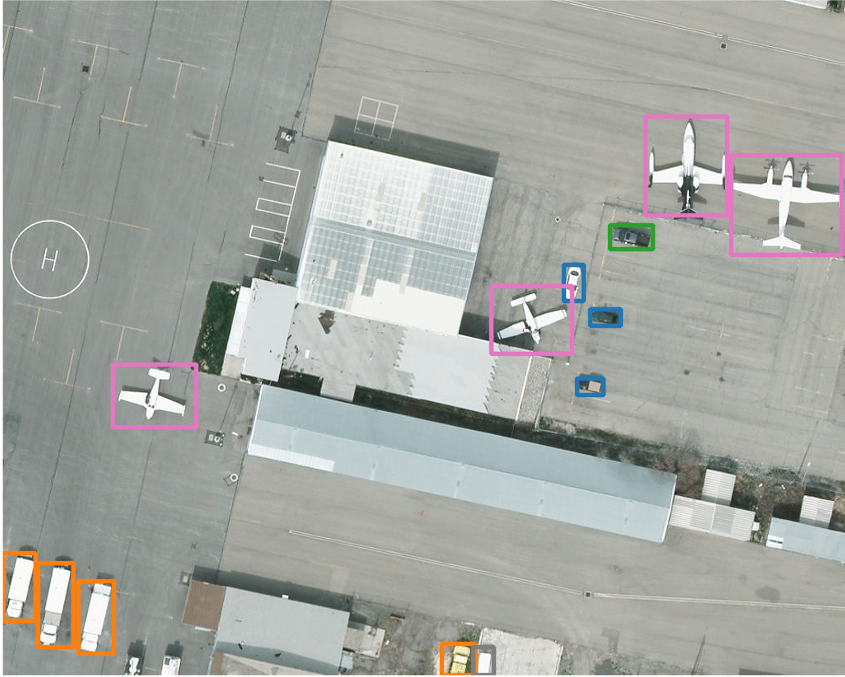}} \hspace{0.05cm}
	{\includegraphics[width=0.46\linewidth]{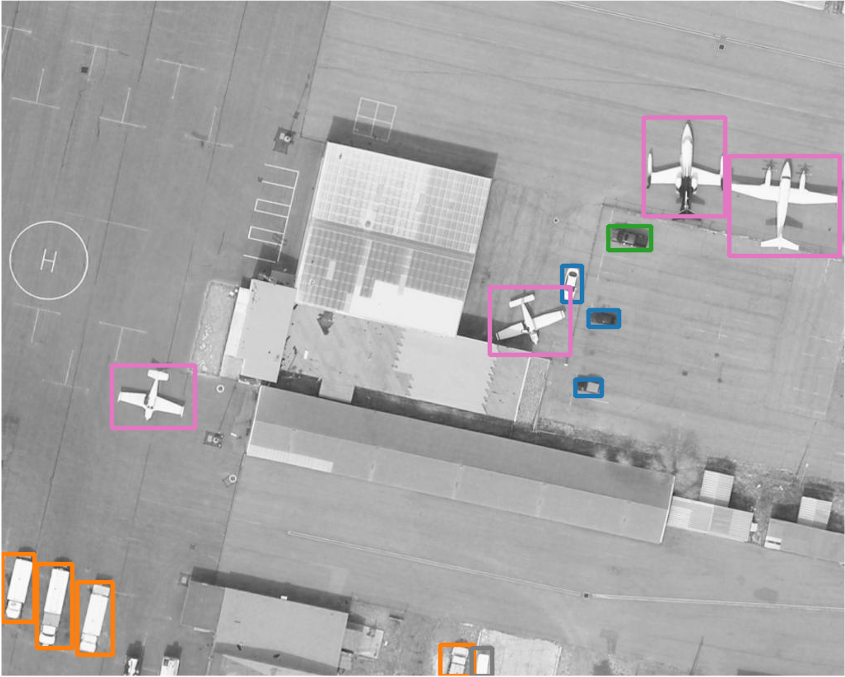}}
	\caption{ 
		Qualitative comparison of multispectral object detection in the VEDAI dataset. 
		The arrangement of the figure is the same as in Fig.~\ref{fig_example_flir}.
		Note that {\color{red} red inverted triangles} indicate FNs, and  {\color{blue} blue inverted triangles}  show FPs.
		Zoom in for more detail.
	}
	\label{fig_example_vedai}
\end{figure}

Additionally, to evaluate the detection results more intuitively, we qualitatively compare the proposed CFT with baselines on FLIR, LLVIP, and VEDAI datasets,  in Fig.~\ref{fig_example_flir}, Fig.~\ref{fig_example_llvip} and 
Fig.~\ref{fig_example_vedai}, respectively.
Visually, even for densely obscured objects, our CFT method can still detect all objects, while the baselines have multiple false positives (FPs) or false negatives (FNs) , i.e., wrong detection.

\subsection{Visual Interpretation}

\begin{figure*}[htbp]
	\centerline{\includegraphics[width=0.8\linewidth]{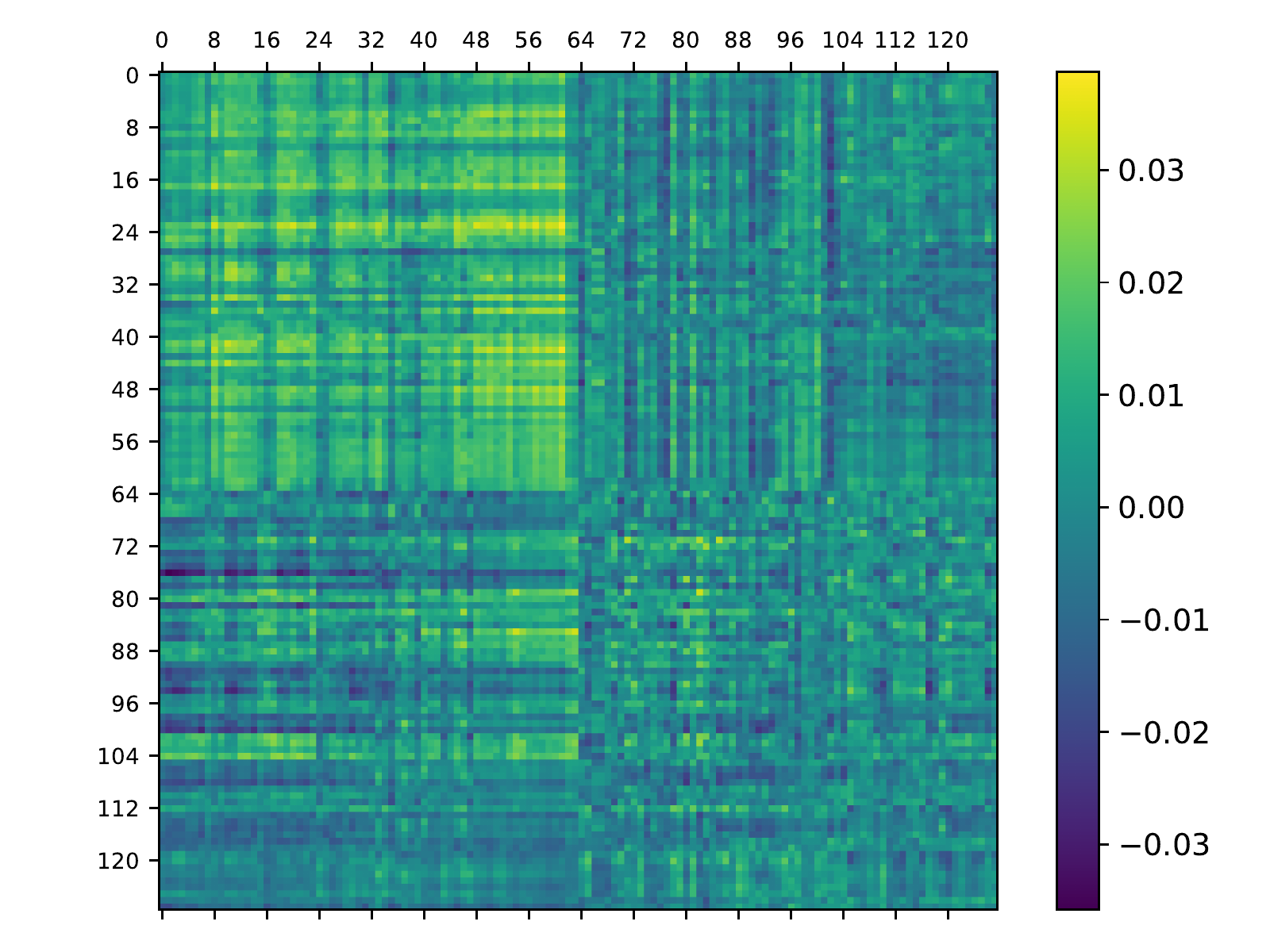}}
	\caption{ Visualization of the correlation matrix . Since $H$=$W$=8 is set in the CFT moudle, the dimension of the correlation matrix is $128\times128$ ($2HW\times2HW$).}
	\label{fig_vis_cor}
\end{figure*}

The correlation matrix $\boldsymbol{\alpha}$ in Eq.~\eqref{eq_alpha}, or more precisely the correlation matrix $\boldsymbol{\alpha}$ of the first CFT in Fig.~\ref{fig_cft}, is visualized in Fig.~\ref{fig_vis_cor} to help clarify clarify the idea described in section "Why Transformer?"
As previously mentioned, the correlation matrix $boldsymbolalpha$ can be broken into four matrix blocks, two of which are inter-modality and two of which are intra-modality.
It is once more confirmed by the depiction in Fig.~\ref{fig_vis_cor}, where the coordinate range of the RGB intra-modality correlation matrix block is $(0,0)$ to $(63,63)$, and the coordinate range of the themral intra-modality correlation matrix block is $(64,64)$ to $(127,127)$, and the rest are two inter-modality correlation matrix blocks.
The illustration in Fig.~\ref{fig_vis_cor} that shows the coordinate range of the RGB intra-modality correlation matrix block as $(0,0)$ to $(63,63)$ and  the themral intra-modality correlation matrix block as $(64,64)$ to $(127,127)$, and the rest as two inter-modality correlation matrix blocks, serves as further confirmation.
Two inter-modality correlation matrix blocks also exhibit some symmetry, as shown in Fig.~\ref{fig_vis_cor}.

\begin{figure*}[htbp]
	\centering 
	\subfigure[Day Sence Input]{
		\begin{minipage}[b]{0.15\linewidth}
			{\includegraphics[width=1\linewidth]{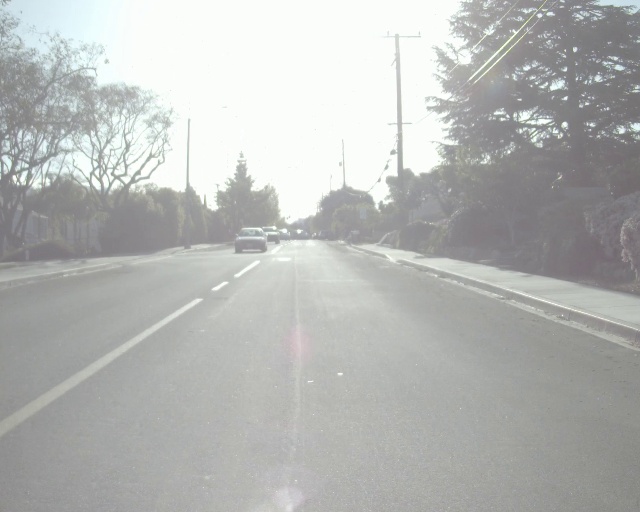}}  \\ \vspace{-0.4cm}
			{\includegraphics[width=1\linewidth]{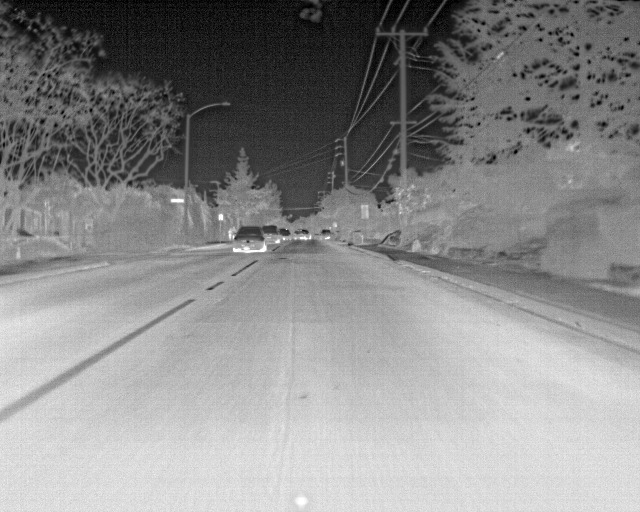}} \\ \vspace{-0.3cm}
			{\includegraphics[width=1\linewidth]{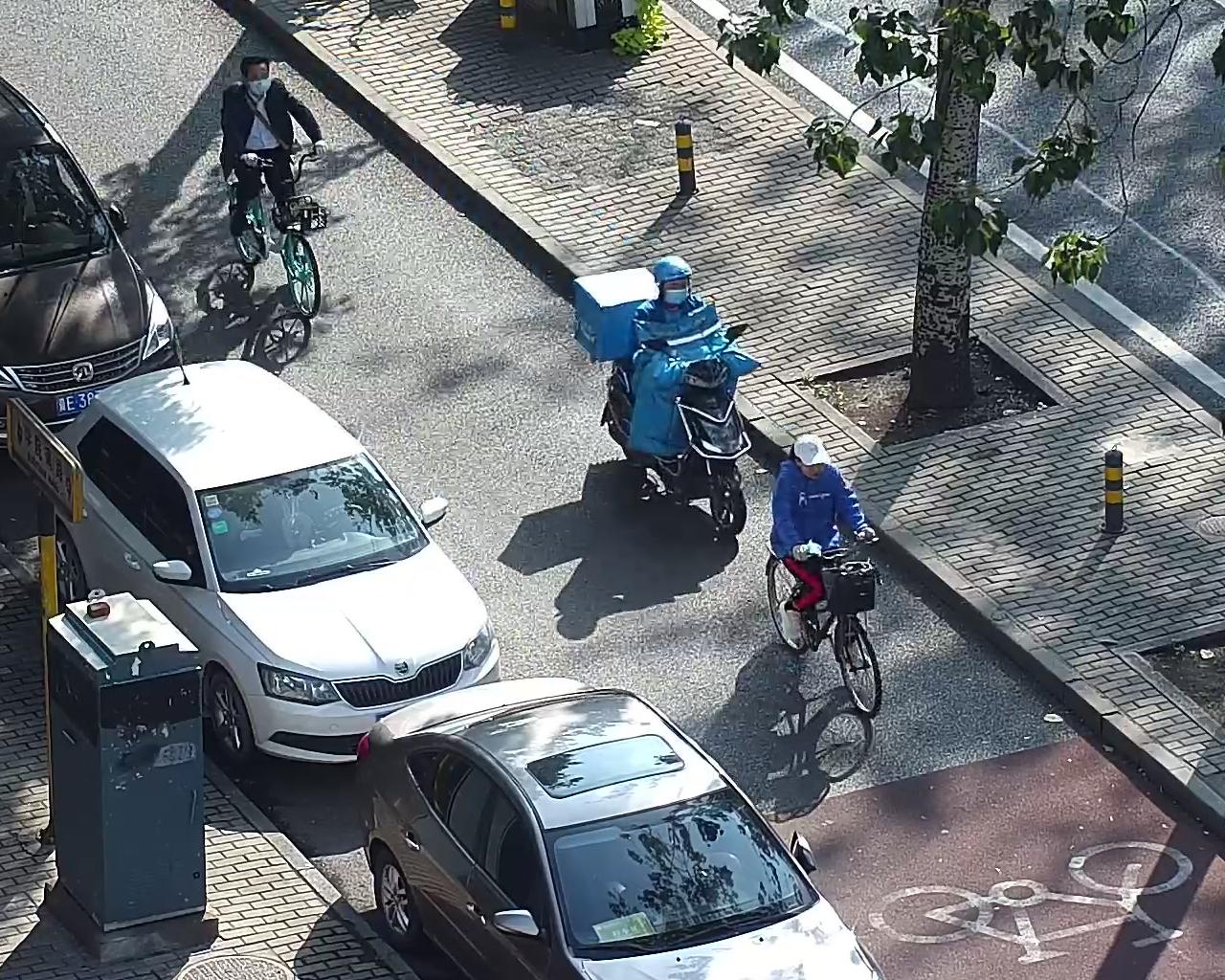}}  \\ \vspace{-0.4cm}
			{\includegraphics[width=1\linewidth]{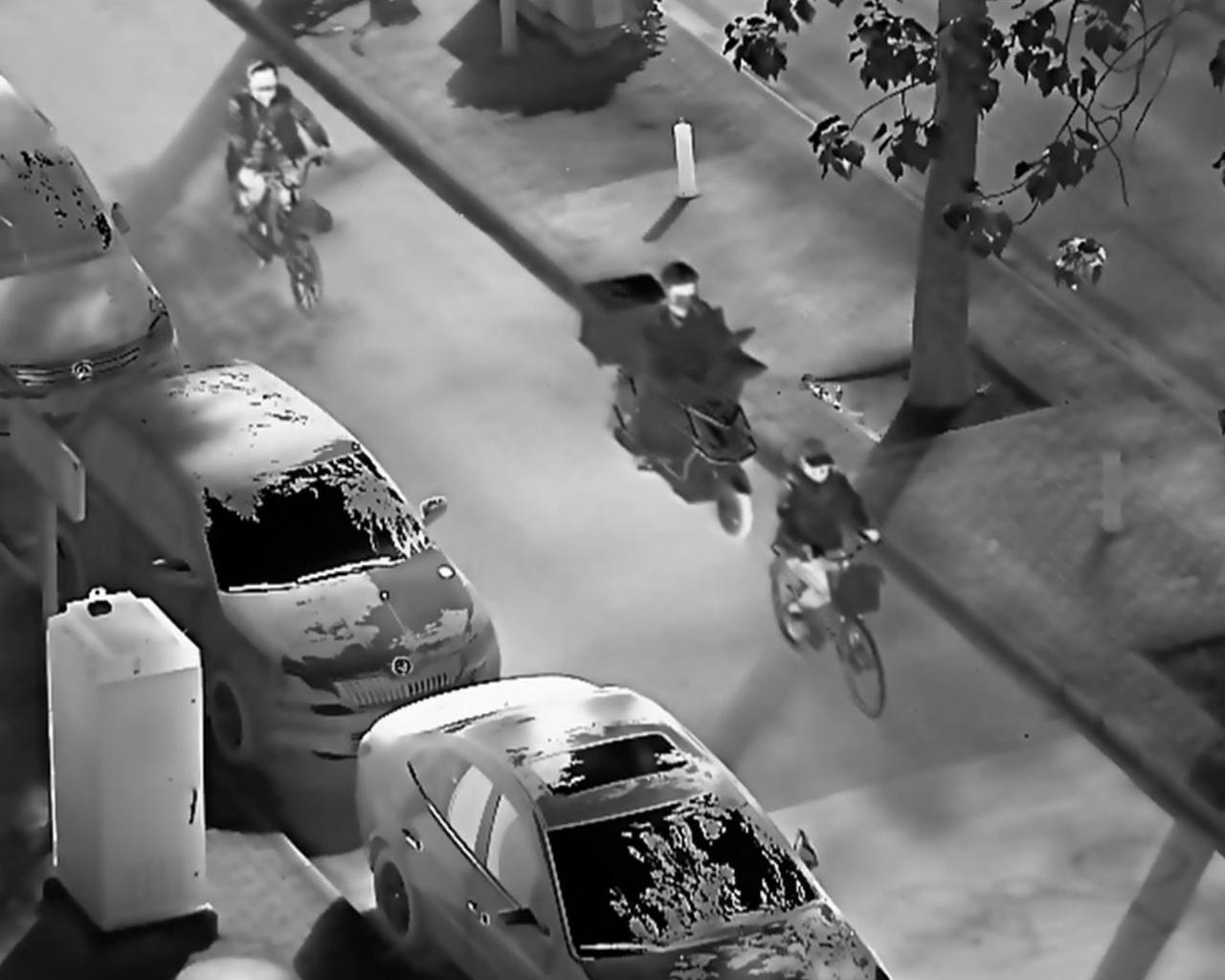}} 
	\end{minipage}}\hspace{-0.1cm}
	\subfigure[Feature]{
		\begin{minipage}[b]{0.15\linewidth}
			{\includegraphics[width=1\linewidth]{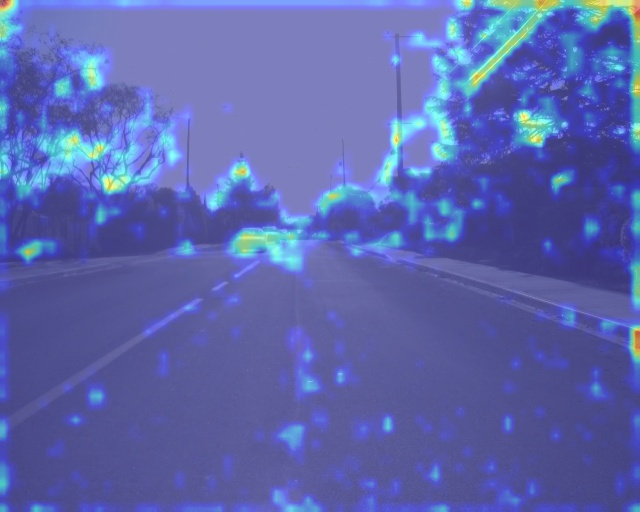}} \\ \vspace{-0.4cm}
			{\includegraphics[width=1\linewidth]{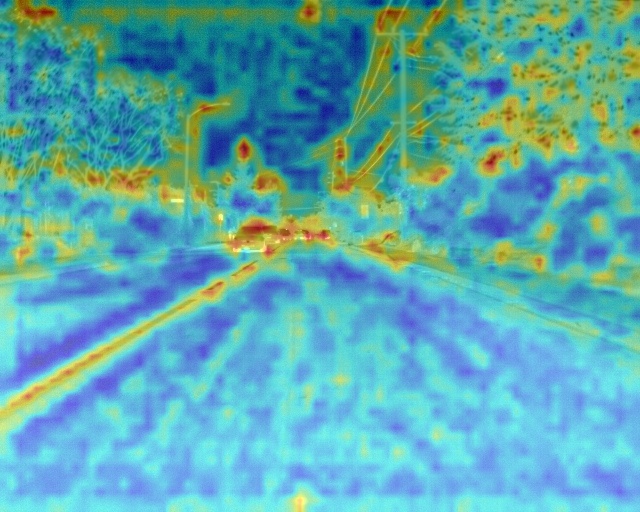}} \\ \vspace{-0.3cm}
			{\includegraphics[width=1\linewidth]{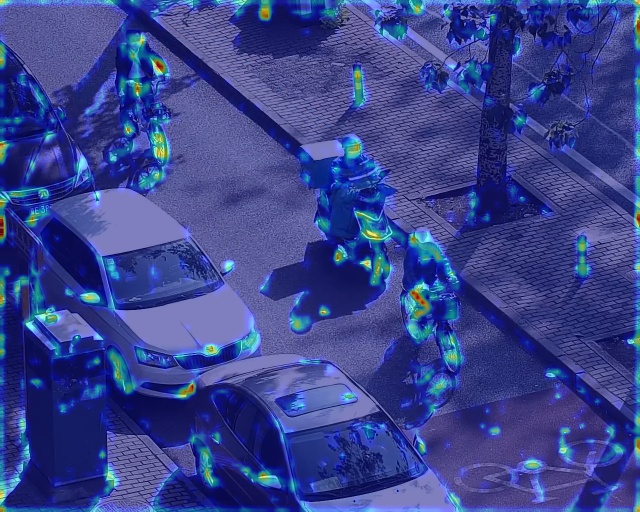}} \\ \vspace{-0.4cm}
			{\includegraphics[width=1\linewidth]{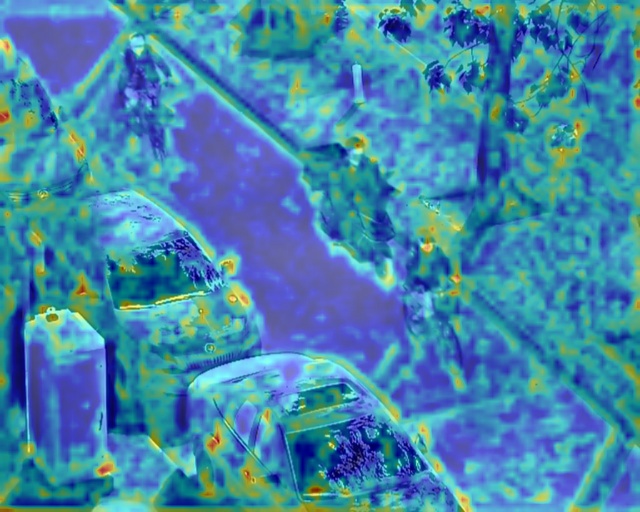}} 
	\end{minipage}}\hspace{-0.1cm}
	\subfigure[After adding CFT]{
		\begin{minipage}[b]{0.15\linewidth}
			{\includegraphics[width=1\linewidth]{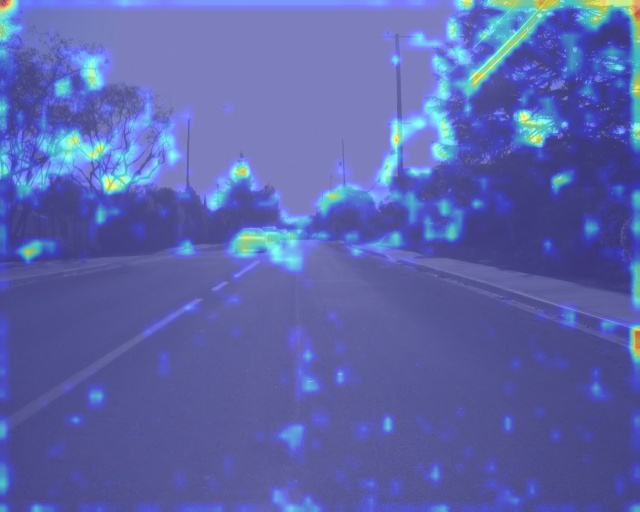}} \\ \vspace{-0.4cm}
			{\includegraphics[width=1\linewidth]{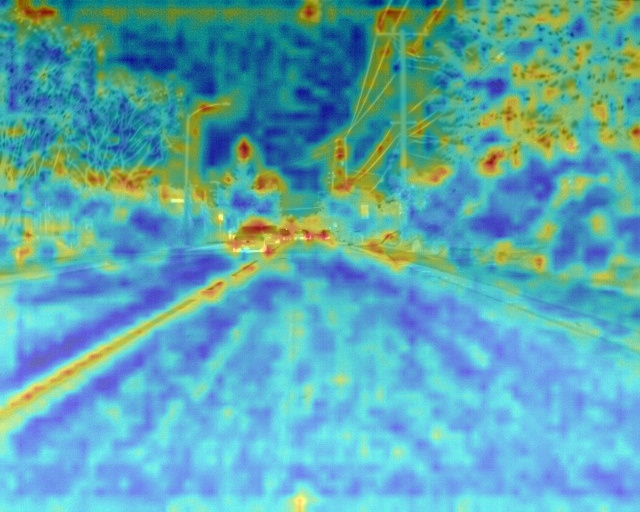}} \\ \vspace{-0.3cm}
			{\includegraphics[width=1\linewidth]{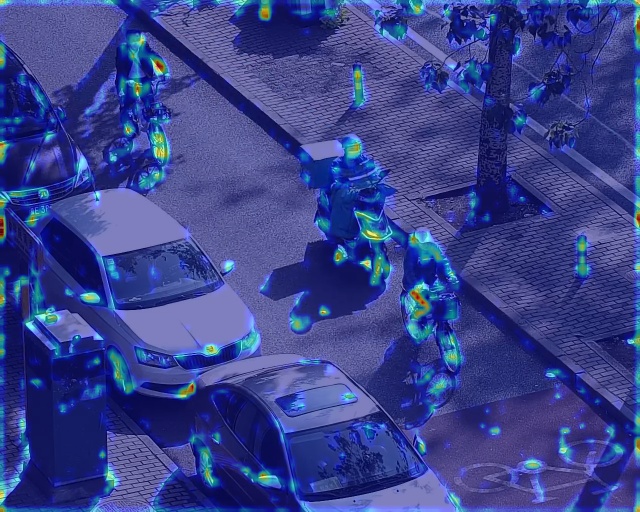}}\\ \vspace{-0.4cm}
			{\includegraphics[width=1\linewidth]{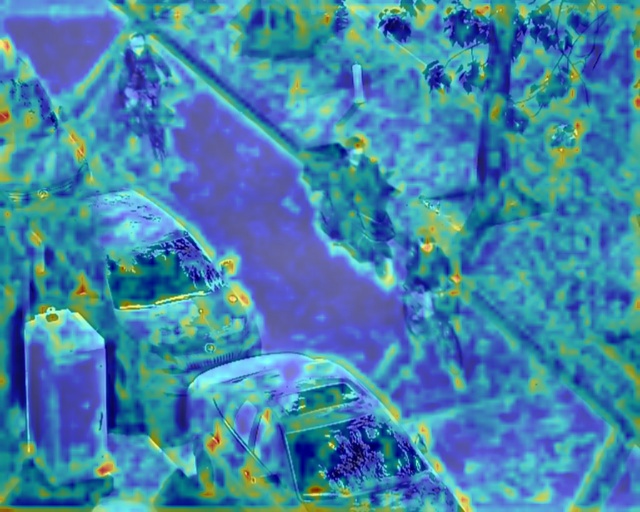}} 
	\end{minipage}}
	\subfigure[Night Sence Input]{
		\begin{minipage}[b]{0.15\linewidth}
			{\includegraphics[width=1\linewidth]{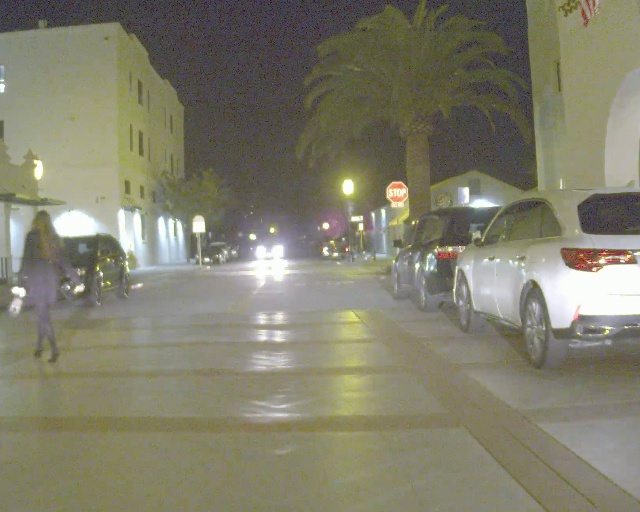}} \\ \vspace{-0.4cm}
			{\includegraphics[width=1\linewidth]{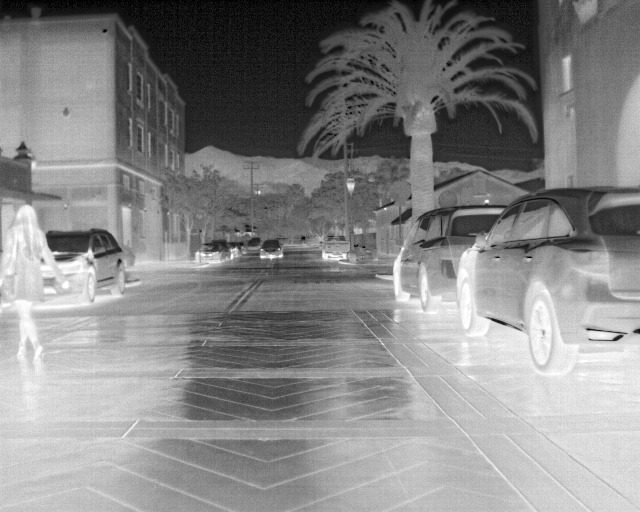}}\\ \vspace{-0.3cm}
			{\includegraphics[width=1\linewidth]{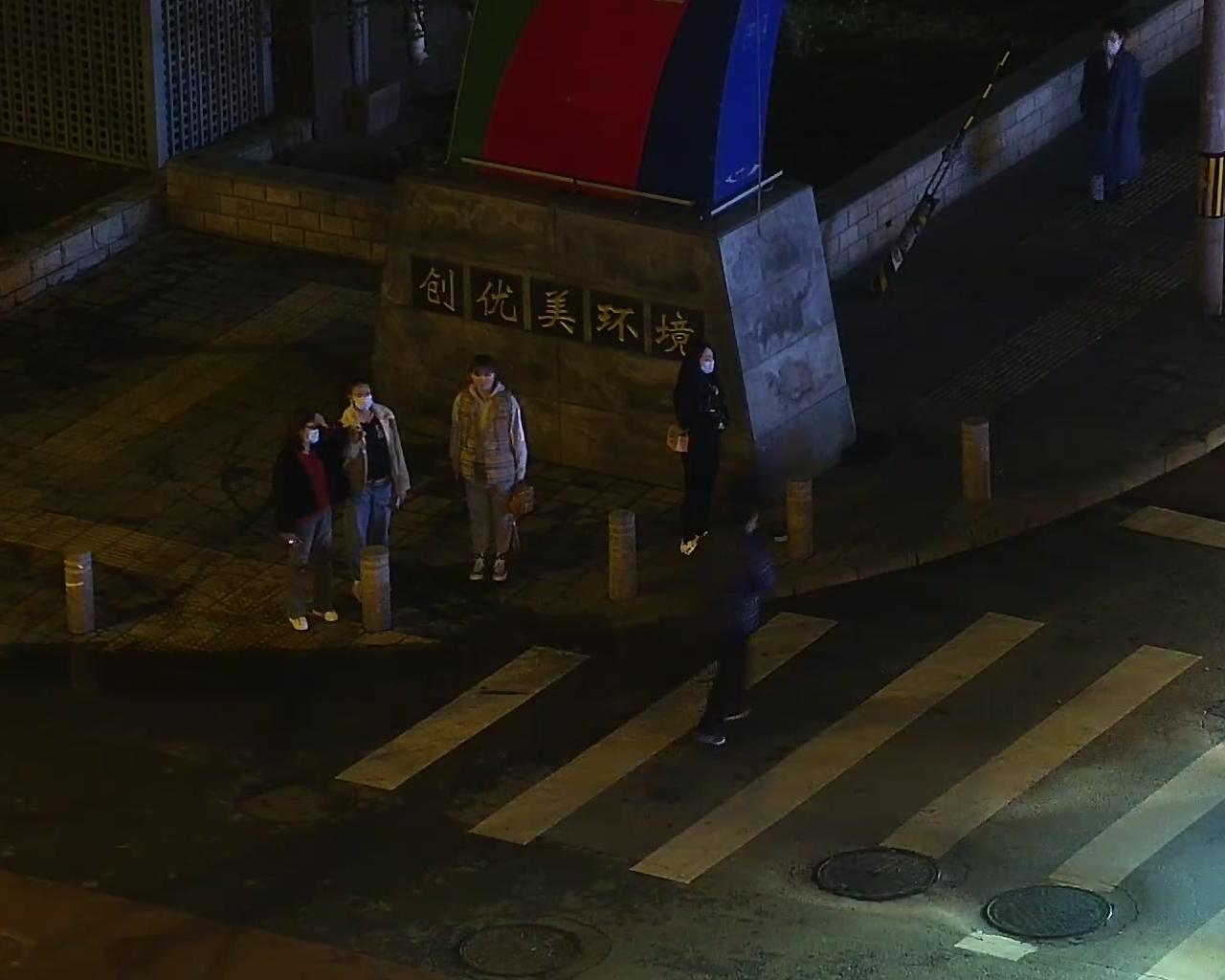}} \\ \vspace{-0.4cm}
			{\includegraphics[width=1\linewidth]{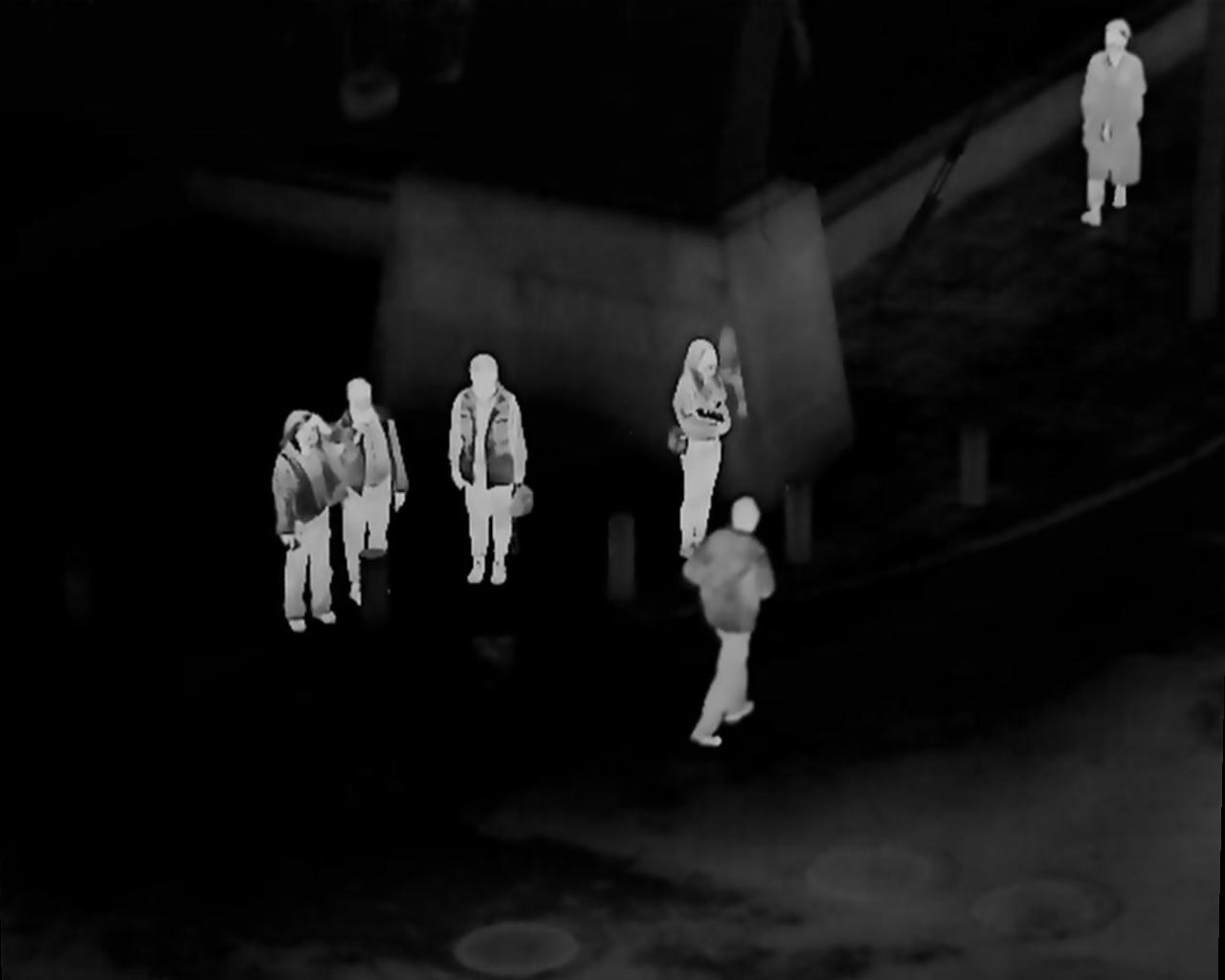}} 
	\end{minipage}}\hspace{-0.1cm}
	\subfigure[Feature]{
		\begin{minipage}[b]{0.15\linewidth}
			{\includegraphics[width=1\linewidth]{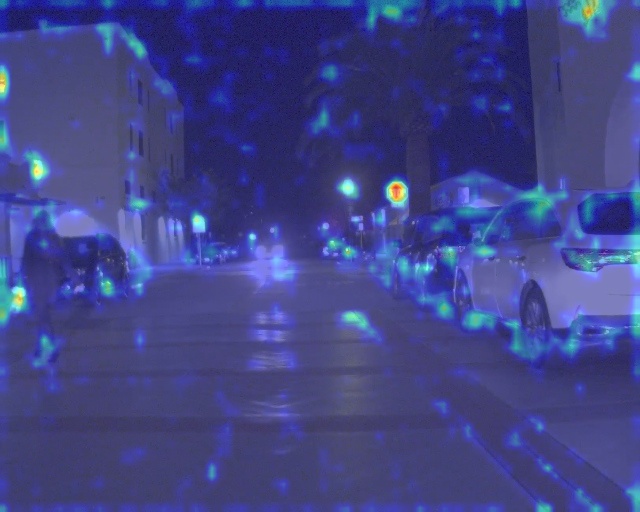}} \\ \vspace{-0.4cm}
			{\includegraphics[width=1\linewidth]{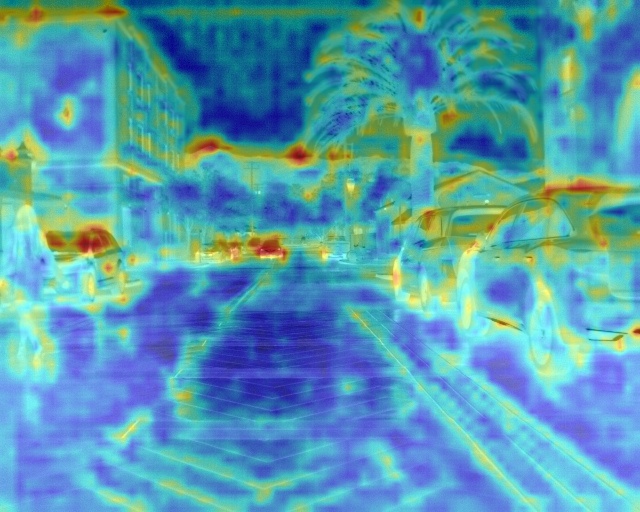}} \\\vspace{-0.3cm}
			{\includegraphics[width=1\linewidth]{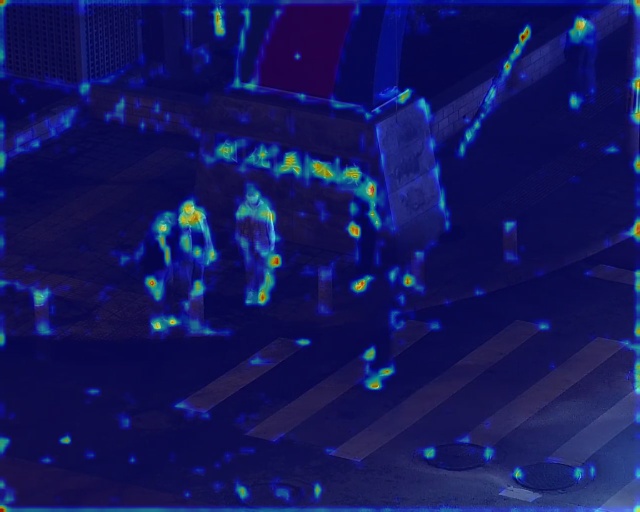}} \\ \vspace{-0.4cm}
			{\includegraphics[width=1\linewidth]{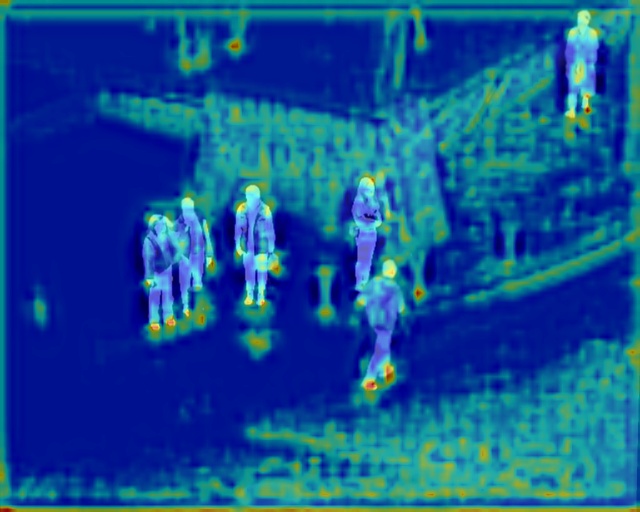}} 
	\end{minipage}}\hspace{-0.1cm}
	\subfigure[After adding CFT]{
		\begin{minipage}[b]{0.15\linewidth}
			{\includegraphics[width=1\linewidth]{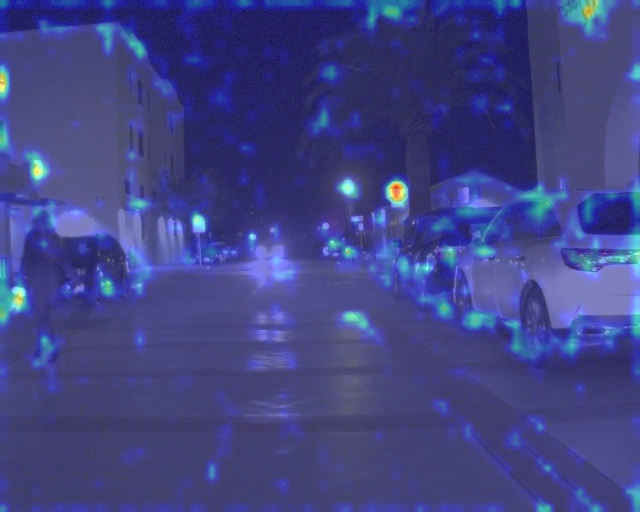}} \\ \vspace{-0.4cm}
			{\includegraphics[width=1\linewidth]{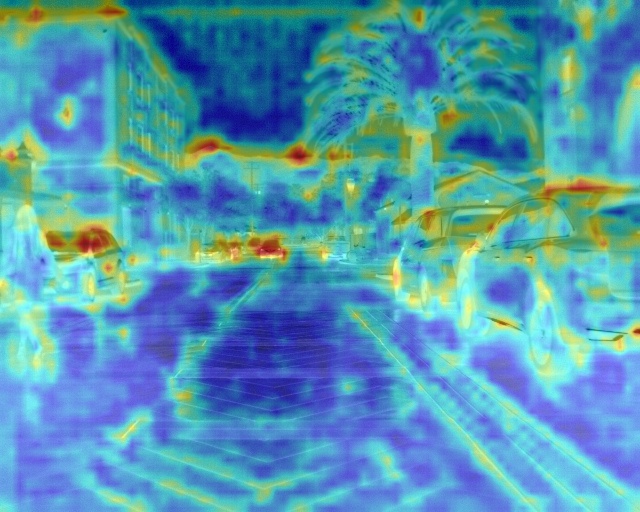}} \\ \vspace{-0.3cm}
			{\includegraphics[width=1\linewidth]{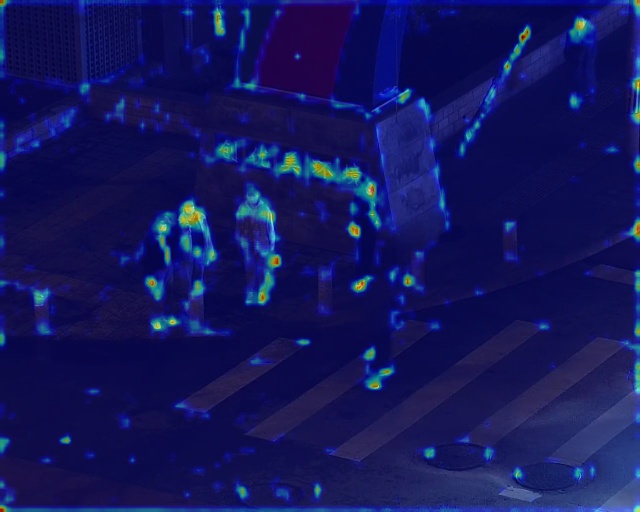}}\\ \vspace{-0.4cm}
			{\includegraphics[width=1\linewidth]{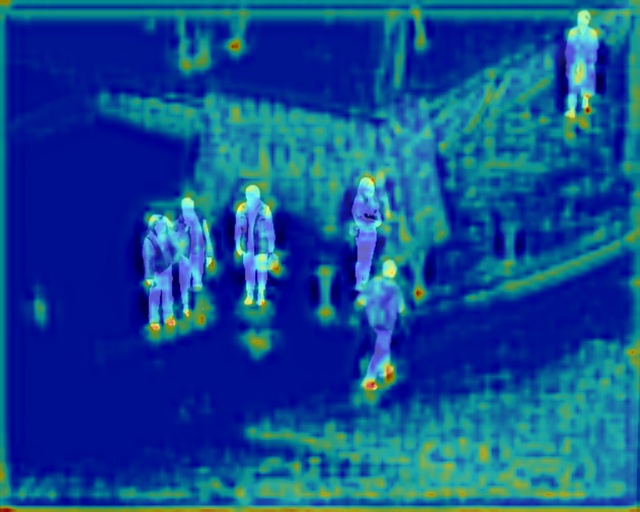}} 
	\end{minipage}}
	
	\caption{
		Four examples of multimodal feature visualization. 
		Two examples on the left are daytime scenes, and  two examples on the right are nighttime scenes.
		The second column and fifth column, subfigure (b) and (e), show feature maps of visual and thermal modalities, as well as the two-branch input for the first CFT module.	
		The third column and last column, subfigure (c) and (f), are the feature maps after adding the CFT module output, that is, $\mathbf{F}_{R3}$ and $\mathbf{F}_{T3}$ in Fig.~\ref{fig_cft}.
		Note the top one means the visual branch, and the bottom one represents the thermal branch in each example.
		Zoom in for more detail.
	}
	\label{fig_fea_vis}
\end{figure*}

Four examples of feature visualization in both daytime and nighttime scenes are shown in Fig.\ref{fig_fea_vis}.
Compared to the thermal modality features, the visual modality features in daytime scene have a stronger focus on the target region of interest and fewer distractions.
In contrast, the visual images are underexposed in the nighttime sense due to the low light, which makes feature extraction faulty by nature.
It is clear from Fig.~\ref{fig_fea_vis}  that the focus of the thermal modality feature is more accurate than the focus of the visual modality feature at night.

The fact that there is almost no difference between the input features and the features after adding the CFT features in Fig.~\ref{fig_cft} is also an intriguing aspect.
The cause of this is that the CFT features are introduced as a residual\cite{HeResnet2015} (as seen in Fig.~\ref{fig_cft}) to enhance the mono-spectral features rather than directly altering the visual or thermal features.

\begin{figure*}[htbp]
	\centering
	\subfigure[Original Feature Visualization]
	{\includegraphics[width=0.47\linewidth]{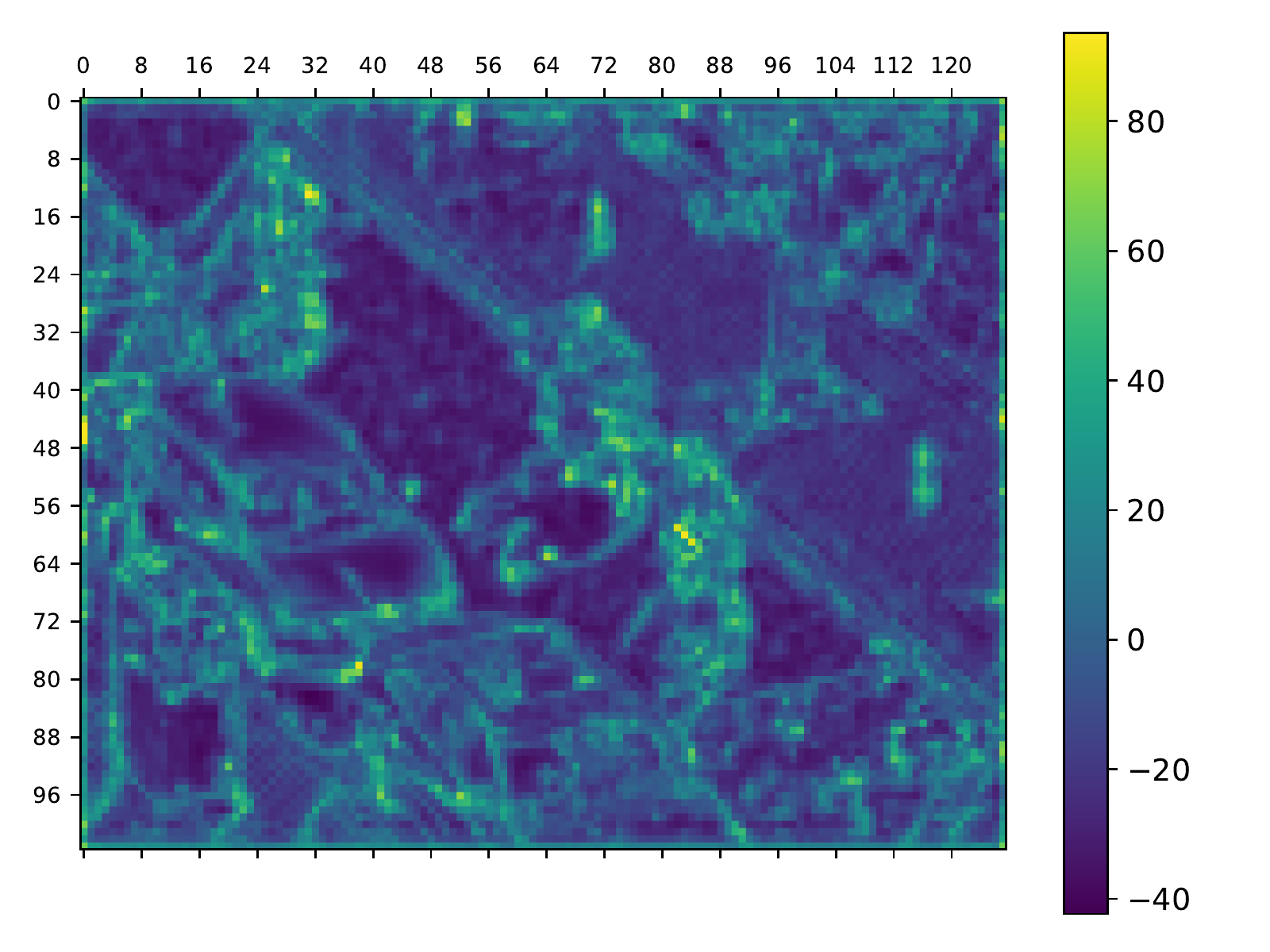}}
	\subfigure[CFT Feature Visualization]
	{\includegraphics[width=0.47\linewidth]{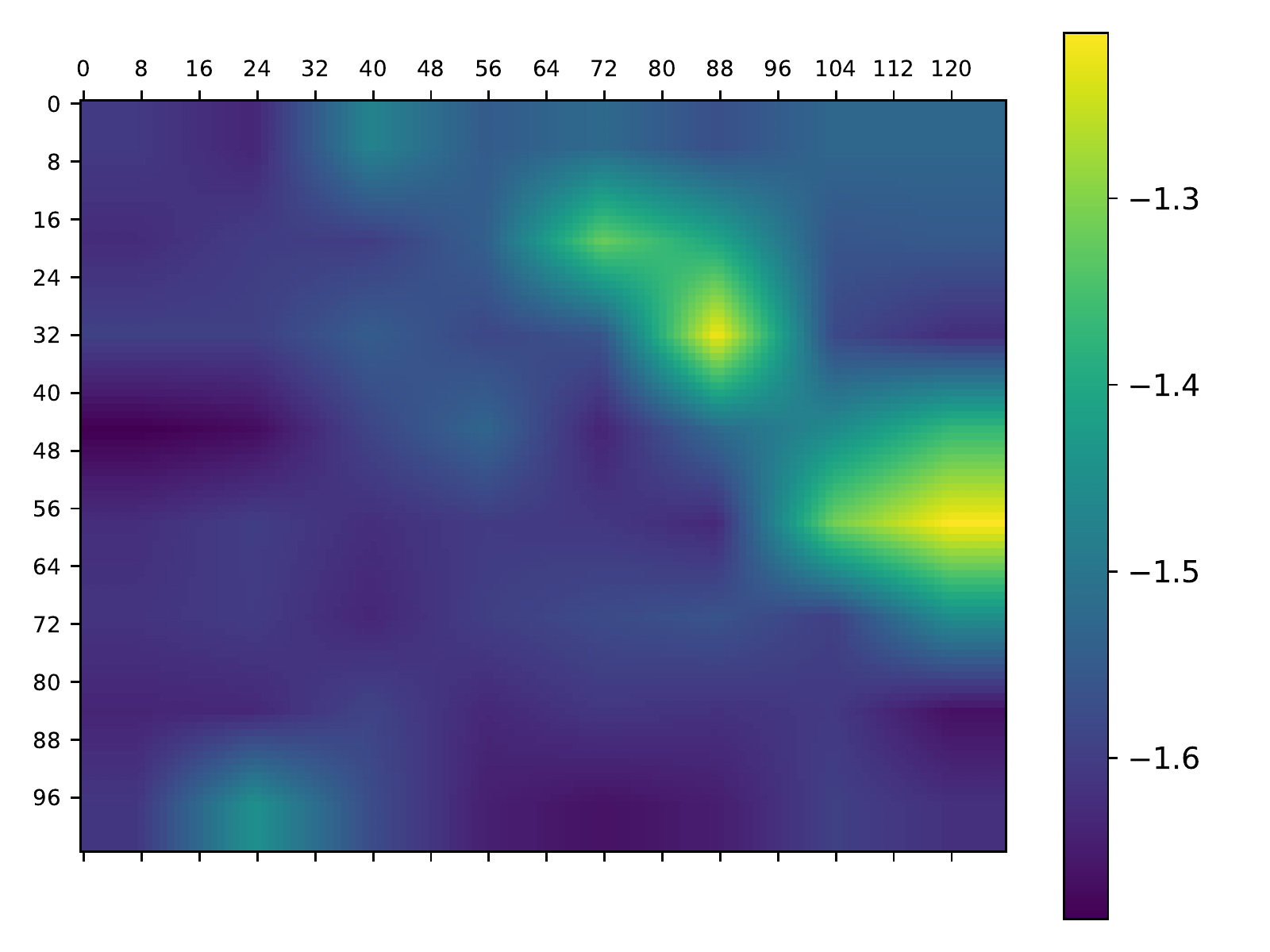}}
		
	\caption{
		Comparison of the original feature visualization and the CFT feature visualization.
		Zoom in for more detail.
	}
	\label{fig_cft_vis}
\end{figure*}

It can be formalized as:
\begin{equation}\label{eq_residual}
	\mathcal{H}(\boldsymbol{x})=\mathcal{F}(\boldsymbol{x})+\boldsymbol{x}
\end{equation}
$\mathcal{H}(\boldsymbol{x})$ forms an identity mapping, i.e., $\mathcal{H}(\boldsymbol{x})=\boldsymbol{x}$, when the CFT module function $\mathcal{F}(\boldsymbol{x})$ equals to 0.
Therefore, the output value of a ideal CFT feature should be as small as possible.
As illustrated in Fig.~\ref{fig_cft_vis}, the value range of the CFT feature visualization is $(-1.7, -1.2)$, whereas the value range of the original feature visualization is $(-45, 100)$.
Clearly, these two value ranges  differ by an order of magnitude.

\subsection{Comparison with State-of-the-art Methods}

{
{\textbf{On FLIR.\footnote{ The latest ranking of this dataset can be checked on Papers with Code (\href{https://paperswithcode.com/sota/multispectral-object-detection-on-flir?}{https://paperswithcode.com/sota/multispectral-object-detection-on-flir?})}}}
Tabel~\ref{tab_flir} compares the results of our approach and other methods. 
It can be observed that our CFT achieves state-of-the-art performance on this dataset.
Furthermore, it has an overwhelming performance improvement, with a minimum difference of 5.8\% and up to 7.5\% on mAP50 between CFT and other multimodal networks .
Even compared to the latest  GAFF \cite{ZhangGuided2021} with ResNet18, our method outperforms 5.8\%, 2.6\%, and 2.7\% on mAP50, mAP75, and mAP respectively.

\begin{table}[htbp]

	\centering
	\small
	\caption{ Comparison of performances on FILR Dataset}
	\setlength{\tabcolsep}{3pt}
	\begin{tabular}{p{2.05cm}p{1.1cm}<{\centering}cp{1.0cm}<{\centering}p{0.9cm}<{\centering}p{0.9cm}<{\centering}}
		\toprule
		Model &Data & Backbone & mAP50  &mAP75 &mAP\\
		\hline
		\multicolumn{6}{c}{mono-modality networks} \\
		\hline
		Faster R-CNN & RGB  & ResNet50 &64.9  & 21.1 &28.9   \\
		Faster R-CNN  & Thermal  & ResNet50 & 74.4 & 32.5 &  37.6 \\
		SSD   & RGB & VGG16 & 52.2  & 15.8  &21.8 \\
		SSD & Thermal & VGG16 &  65.5&22.4  & 29.6 \\
		YOLOv3& RGB  & Darknet53 & 58.3 & 19.8 &  25.7 \\
		YOLOv3& Thermal  & Darknet53 & 73.6 & 31.3 & 36.8 \\
		YOLOv5 & RGB  &CSPD53 & 67.8 & 25.9 & 31.8 \\
		YOLOv5 & Thermal &CSPD53 & 73.9&\textbf{35.7} &39.5\\
		\hline
		\multicolumn{6}{c}{multi-modality networks} \\
		\hline
		Halfway  \cite{ZhangMultispectral2020} &RGB+T &VGG16& 71.2 &- &- \\
		CFR\_3 \cite{ZhangMultispectral2020}&RGB+T &VGG16 & 72.4  & - &- \\ 
		GAFF \cite{ZhangGuided2021} &RGB+T &ResNet18 & 72.9 &{32.9} & 37.5\\
		GAFF \cite{ZhangGuided2021} &RGB+T &VGG16 & 72.7 & 30.9 & 37.3\\		
		Baseline(Ours) &RGB+T & CSPD53 & {73.0} & {32.0} & {37.4} \\
		CFT(Ours) &RGB+T &CFB & \textbf{78.7} & {35.5} &\textbf{40.2} \\
		\bottomrule
	\end{tabular}
	\label{tab_flir}
\end{table}

{\textbf{On LLVIP.}\footnote{ Check the ranking website: \href{https://paperswithcode.com/sota/pedestrian-detection-on-llvip?}{https://paperswithcode.com/sota/pedestrian-detection-on-llvip?}} }
Table~\ref{tab_llvip} presents the detection performance of CFT and other mono-modality networks (especially YOLOv5 which is the foundation of our algorithm), on the LLVIP dataset.
It indicates that a more accurate detection (mAP50:97.5, mAP75:72.9, mAP:63.6) can be carried out by interacting and fusing the complementary features of different modalities based on our CFT module.
\begin{table}[htbp]
	
	\centering
	\caption{Comparison of performances on LLVIP Dataset}
	\setlength{\tabcolsep}{3pt}
	\begin{tabular}{p{2.05cm}p{1.1cm}<{\centering}cp{1.0cm}<{\centering}p{0.9cm}<{\centering}p{0.9cm}<{\centering}}
		\toprule
		Model & Data & Backbone& mAP50  & mAP75 & mAP\\
		\hline
		\multicolumn{6}{c}{mono-modality networks} \\
		\hline
		Faster R-CNN  & RGB  & ResNet50 &91.4  &48.0  &49.2   \\
		Faster R-CNN & Thermal  & ResNet50 &  96.1 &68.5   &61.1   \\
		SSD  & RGB & VGG16 &  82.6  &31.8  &39.8 \\
		SSD  & Thermal & VGG16 & 90.2 &57.9  &53.5  \\
		YOLOv3 & RGB &Darknet53 & 85.9 &37.9& 43.3\\
		YOLOv3 & Thermal & Darknet53&89.7& 53.4& 52.8\\
		YOLOv5 & RGB  &CSPD                                                                                                                                                                                                                                                                                                                                                                                                                                              53 &90.8 & 51.9 & 50.0\\
		YOLOv5 & Thermal &CSPD53&94.6 & {72.2 } &61.9\\
		\hline
		\multicolumn{6}{c}{multi-modality networks} \\
		\hline
		Baseline(Ours) &RGB+T & CSPD53 & {95.8} & {71.4} & {62.3} \\
		CFT(Ours) & RGB+T& CFB&\textbf{97.5} & \textbf{72.9}  &\textbf{63.6} \\
		\bottomrule
	\end{tabular}
	\label{tab_llvip}
\end{table}

{\textbf{On VEDAI.} }
Tabel~\ref{tab_vedai} reports the vehicle detection performance of our and other methods on the multispectral aerial imagery dataset. 
Again, we can observe that our CFT is more accurate than the best mono-modality network YOLO-fine in mAP50 (9.3\%). 
For multi-modality networks, the detection performance of our baseline surpasses the previous network (YOLOv3 with mid-level fusion) in the mAP (2.2\%), while our CFT method does better. 
It provides non-negligible gains for all the considered evaluation metrics (mAP50:{\color{blue}$\uparrow$9.3}, mAP:{\color{blue}$\uparrow$10.0}).

\begin{table}[htbp]
	
	\centering
	\caption{Comparison of performances on VEDAI Dataset}
	\setlength{\tabcolsep}{3pt}
	\begin{tabular}{lcccc}
		\toprule
		Model & Data & Backbone  & mAP50 & mAP\\
		\hline
		\multicolumn{5}{c}{mono-modality networks} \\
		\hline
		Faster R-CNN  & RGB & ResNet50 & 63.5  & 39.4	 \\
		Faster R-CNN & Thermal & ResNet50 & 72.2   & 42.6	 \\
		SSD   & RGB & VGG16 & 70.9 & - \\
		SSD  & Thermal & VGG16 & 69.8 & - \\
		SSSDET \cite{MandalSSSDET2019} & RGB & shallow  network & - & 46.0 \\
		YOLO-fine \cite{PhamYOLO-Fine2020} & RGB &  Darknet53 & 76.0 & - \\
		YOLO-fine \cite{PhamYOLO-Fine2020}   & Thermal &  Darknet53 &   75.2& - \\
		YOLOv3 & RGB  & Darknet53 & 70.0  &  42.0 \\
		YOLOv3 & Thermal  & Darknet53 & 69.3  & 43.3 \\
		YOLOv5 & RGB &  CSPD53 & 74.3 & 46.2 \\
		YOLOv5  & Thermal &  CSPD53 &   74.0 & 46.1\\
		\hline
		\multicolumn{5}{c}{multi-modality networks} \\
		\hline
		early fusion \cite{DhanarajVehicle2020} & RGB+T &  Darknet53 & - & 44.0  \\ 
		Mid fusion \cite{DhanarajVehicle2020}  & RGB+T & Darknet53 & - & 44.6  \\ 
		Baseline(ours) & RGB+T & CSPD53 &79.7  & 46.8  \\
		CFT(ours) & RGB+T & CFB &\textbf{85.3}  & \textbf{56.0}  \\
		\bottomrule
	\end{tabular}
	\label{tab_vedai}
\end{table}

}

\section{Conclusion}

In this work, we propose a novel Transformer-based fusion approach, namely Cross-Modality Fusion Transformer (CFT), to
learn long-range dependencies and integrate global contextual information, thereby the representation power of two-stream CNNs is enhanced in multispectral object detection. 
More specifically, the CFT modules are densely inserted into the backbone to integrate features, hence the inherent complementarity between different modalities can be fully exploited.
{Moreover, we show the fusion process of the multi-modality feature maps by the CFT module from both formula and implementation points of view.
The proposed method achieves state-of-the-art performances of 78.5, 97.5, and 85.3 mAP50 on FLIR, LLVIP, and VEDAI datasets, respectively.  
To present the general effectiveness, the proposed CFT module is combined with three classical detectors, i.e., YOLOV5, YOLOv3, and Faster R-CNN.
The experimental results show that the proposed CFT improves the performance of multispectral object detection with either the one-stage or two-stage detector by a clear margin.}
Since our approach is simple yet effective, it may be applied to other computer vision fields such as RGB-LiDAR, RGB-D, stereo image SR tasks, etc.

\section{Acknowledgement}

This work is supported by the National Natural Science Foundation of China under Grant No.U20B2056 and No.11872034.

{
	\bibliography{mybibfile.bib}
}

\end{document}